\documentclass[sn-mathphys,iicol]{sn-jnl}

\jyear{2022}%

\usepackage{combelow}

\usepackage[utf8]{inputenc}
\usepackage{bm}
\usepackage{bbold}
\usepackage{mathtools}

\usepackage{graphicx}
\usepackage{subfigure}
\usepackage{scalerel,amssymb}
\DeclareMathAlphabet{\pazocal}{OMS}{zplm}{m}{n}

\usepackage{braket}
\usepackage{tensor}
\usepackage{slashed}
\usepackage{xcolor}
\usepackage{ulem}
\usepackage{xr}

\definecolor{purple}{rgb}{0.8,0,0.6}

\def\bp{{\mathbf{p}}}
\def\bk{{\mathbf{k}}}

\usepackage{fancyhdr}
\usepackage{lastpage}

\newcommand{\beqn}{\begin{eqnarray}}
\newcommand{\eeqn}{\end{eqnarray}}
\newcommand{\beqs}{\begin{subequations}}
\newcommand{\eeqs}{\end{subequations}\\[-2mm]\noindent}
\newcommand{\eq}[1]{(\ref{#1})}
\newcommand{\bs}{\boldsymbol}

\newcommand{\cS}{{\mathcal S}}
\newcommand{\cL}{{\mathcal L}}
\newcommand{\cH}{{\widehat{H}}}
\newcommand{\HVW}{{\mbox{\tiny{HVW}}}}
\newcommand{\CVW}{{\mbox{\tiny{CVW}}}}

\begin{document}

\title{Vortical effects in Dirac fluids with vector, chiral and helical charges}

\author[1]{\fnm{Victor E.} \sur{Ambru\cb{s}}}\email{victor.ambrus@e-uvt.ro}

\author*[2]{\fnm{M. N.} \sur{Chernodub}}\email{maxim.chernodub@univ-tours.fr}

\affil*[1]{
\orgdiv{Department of Physics}, 
\orgname{West University of Timi\cb{s}oara},
\orgaddress{\street{Bd.~Vasile P\^arvan 4}, \city{Timi\cb{s}oara}, \postcode{300223}, \country{Romania}}}

\affil[2]{
\orgdiv{Institut Denis Poisson, CNRS UMR 7013},
\orgname{Universit\'e de Tours}, 
\orgaddress{\city{Tours}, \postcode{37200}, \country{France}}}

\abstract{Helicity of free massless Dirac fermions is a conserved, Lorentz-invariant quantity at the level of the classical equations of motion. For a generic ensemble consisting of particles and antiparticles, the helical and chiral charges are different conserved quantities. The flow of helicity can be modelled by the helicity current, which is again conserved in the absence of interactions. Similar to the axial vortical effect which generates an axial (chiral) current, the helicity current is induced by vorticity in a finite temperature medium with vector (electrical) charge imbalance via the helical vortical effects, leading to new nondissipative transport phenomena. These phenomena lead to the appearance of a new 
hydrodynamic excitation, the helical vortical wave. Our results suggest the existence of a new type of triangle anomalies in QED which involve the helicity currents in addition to the standard vector and axial currents. Further exploiting the conservation of the helical current, we show that a finite helical chemical potential may be used to characterise thermodynamic ensembles of fermions similarly 
to, but independently of, the vector charge and chirality. 
We derive the pressure $P$ for fermions at finite vector, axial and helical chemical potentials and show that the quantities arising in anomalous transport, including various vortical and circular conductivities and the shear-stress coefficients, can be obtained by differentiation of $P$ with respect to the appropriate chemical potentials. Finally, we calculate the helicity relaxation time in the quark-gluon plasma above the crossover and show that it is similar to that for the axial charge.}

\keywords{Helicity, Vorticity, Dirac fermions, Vortical effects}



\maketitle

\section{Introduction}\label{sec:intro}

Massless or nearly-massless fermions appear in many areas of physics, including theories of fundamental interactions, cosmological models of the early Universe, ultra-hot relativistic plasmas, and superfluids~\cite{volovik2009universe}. Many relativistic phenomena are now available for experimental verification in recently discovered crystals of Dirac and Weyl semimetals, where the massless fermions appear as quasiparticle excitations~\cite{DiracWeylReview}. The most important properties of these excitations are usually associated with their vector (gauge) and axial (chiral) symmetries that affect, in the case of semimetals,  electromagnetic~\cite{Vazifeh_2013}, thermal~\cite{gooth2017experimental}, and elastic~\cite{Cortijo_2015} responses of these materials. Many of the unusual features of these semimetals are associated with the quantum anomaly that breaks the continuous axial symmetry of an underlying classical theory~\cite{FujikawaBook}. Similar anomalies lead to exotic transport phenomena of quarks, mediated by the topology of evolving gluon fields in expanding quark-gluon plasma (QGP) of heavy-ion collisions (HIC)~\cite{Kharzeev:2013ffa}.

In addition to the vector and axial charges, massless fermions can be characterised via a third, well-known, and, simultaneously, often-forgotten quantity: the helicity. The fermionic helicity is sometimes confused with the chirality, even though these quantities reflect different physical properties of fermions~\cite{Pal:2010ih}. To highlight the importance of helicity, we demonstrate the existence of a set of new transport phenomena emerging in a gas of rotating massless fermions, the Helical Vortical Effects (HVEs), that differ substantially from their chiral counterparts, the Chiral Vortical Effects (CVEs)~\cite{CVE:1,CVE:2,CVE:3,CVE:4,footnote:0,Yamamoto:2015gzz}. 

Hydrodynamics of relativistic plasmas with nonzero vorticity has attracted significant attention~\cite{Jiang:2015cva,Boyarsky:2015faa,Pavlovic:2016gac,Hattori:2017usa,Rybalka:2018uzh}. The HVEs may see their applications in noncentral ultrarelativistic 
HIC that create a nearly perfect fluid of QGP,
the most vortical fluid ever known~\cite{STAR:2017ckg}. 
Exploiting the conservation of the helicity current $J^\mu_H$ for 
free fermions, we model helicity imbalance by means of the helicity 
chemical potential $\mu_H$, alongside the more familiar 
vector and axial chemical potentials, $\mu_V$ and $\mu_A$, respectively.
At a classical (non-quantum) level, we consider an ensemble 
of fermions at finite temperature and chemical potentials $\mu_\ell$ 
distributed according to the Fermi-Dirac distribution. Within this 
thermodynamic framework, we take derivatives of the pressure $P$ 
with respect to $\mu_A$ and show that the resulting quantities can be
related to the vortical transport coefficients (e.g., the vortical and
circular conductivities $\sigma^\omega_\ell$ and $\sigma^\tau_\ell$, respectively) 
appearing in anomalous transport. We then employ finite-temperature 
quantum field theory to compute the expectation values of the charge 
currents $J^\mu_\ell$ of vector ($\ell = V$), axial ($\ell = A$), and helical ($\ell = H$) charges and of the stress-energy tensor $T^{\mu\nu}$ at finite temperature $T$,
finite chemical potentials (including finite $\mu_H$)
and finite rotation parameter $\Omega$.
We show that all ``classical'' quantities receive quantum corrections that 
are of higher order with respect to the angular frequency~$\Omega$.

This paper is organised as follows. In Sec.~\ref{sec:helical}, the 
vector ($J^\mu_V$), axial ($J^\mu_A$) and helical ($J^\mu_H$) charge 
currents are introduced as a triad 
corresponding to three Abelian symmetry groups of the classical Dirac 
Lagrangian. Also here, their operator form after second 
quantisation and their properties under the C, P, and T 
transformations are derived and the 
chemical potentials $\mu_{V/A/H}$
associated to the conserved charge operators $\widehat{Q}_{V/A/H}$
are introduced.
Section~\ref{sec:RKT} introduces the classical description 
of fermions at finite $T$ and $\mu_\ell$, based on the Fermi-Dirac 
distribution. 
The properties of thermal states under rigid rotation at finite chemical 
potentials are discussed in Sec.~\ref{sec:QFT}. The full details of 
the computation are presented in Appendix~\ref{app:comp}.
We comment on the possible implications for new types of triangle anomalies
involving the helical vertex ({\it Helical triangle anomalies}), the helicity relaxation time in the strongly-interacting quark-gluon plasma,
new wave-like excitations in the helical-vortical matter 
({\it Helical vortical waves}) and new vortices in Dirac fluids
in Section~\ref{sec:app}. 
Section~\ref{sec:conc} concludes this paper.

Throughout this paper, Planck units ($\hbar = c = k_B = 1$) are used. 
We use the convention $\varepsilon^{0123} = (-g)^{-1/2}$ 
for the Levi-Civita tensor and the $(+---)$ metric signature.

\section{Vector-helical-axial triad for massless Dirac fermions}\label{sec:helical}

In this section, we review the vector and axial currents 
$J^\mu_V$ and $J^\mu_A$ from the perspective of the 
invariance of the free theory under the abelian $U(1)_V$ 
and $U(1)_A$ symmetries. On the same footing, we introduce 
in Subsec.~\ref{sec:helical:transf}
the helicity current based on the invariance of the free theory 
under the abelian and non-local helical transformations group 
$U(1)_H$. The covariance properties of the helicity current $J^\mu_H$ under Lorentz transformations are discussed in Subsec.~\ref{sec:helical:cov}. In Subsec.~\ref{sec:helical:Q}, we discuss the 
charge operators $\widehat{Q}_{V/A/H}$ corresponding to the vector, 
axial and helical charges, obtained after 
second quantisation. The CPT parities of the corresponding 
charge current operators $\widehat{J}^\mu_{V/A/H}$ are considered 
in Subsec.~\ref{sec:helical:CPT}. Finally, in 
Subsec.~\ref{sec:helical:muH} we discuss the thermodynamic 
interpretation of the helicity chemical potential $\mu_H$ from 
the perspective of its associated charge operator, $\widehat{Q}_H$.

\subsection{$U(1)_H$ group of helical transformations and helicity 
current}\label{sec:helical:transf}

We consider one species of free massless Dirac fermions in a flat $(3+1)$d Minkowski spacetime, with the Lagrangian
\begin{equation}
 \cL = \frac{i}{2} (\overline{\psi} \gamma^\mu \partial_\mu \psi - 
 \partial_\mu \overline{\psi} \gamma^\mu \psi),
 \label{eq:L}
\end{equation}
where $\overline{\psi} = \psi^\dagger \gamma^0$ is the Dirac adjoint of the 4-spinor $\psi$. 
We take the $4\times 4$ gamma matrices $\gamma^\mu$ ($\mu = 0, \dots 3$) in the Dirac
representation.

The classical Dirac Lagrangian~\eq{eq:L} is known to be
invariant under the vector $U(1)_V$ global Abelian symmetry: 
\begin{align}
U(1)_V:& & 
\psi \rightarrow& e^{i\alpha_V} \psi, &
{\bar\psi} \rightarrow& e^{-i\alpha_V} {\bar \psi}, \label{eq:U1:V}
\end{align}
which allows the vector current to be derived via Noether's theorem, 
\begin{equation}
 J^\mu_V = \bar{\psi} \gamma^\mu \psi,\label{eq:JV}
\end{equation}
satisfying $\partial_\mu J^\mu_V = 0$ when $\psi$ obeys the 
Dirac equation,
\begin{equation}
 \slashed{\partial} \psi = 0,\label{eq:diraceq}
\end{equation}
where $\slashed{\partial} = \gamma^\mu \partial_\mu$ is the Feynman 
slash notation.

Another well-known symmetry is related to the $U(1)_A$ symmetry of
the Lagrangian \eqref{eq:L} under chiral transformations,
\begin{align}
 U(1)_A:& &
 \psi \rightarrow& e^{i\alpha_A \gamma^5} \psi, &
 {\bar\psi} \rightarrow& {\bar \psi} e^{i\alpha_A \gamma^5}, \label{eq:U1:A}
\end{align}
where $\gamma^5 = i \gamma^0 \gamma^1 \gamma^2 \gamma^3$ is the 
fifth gamma matrix. The $U(1)_A$ symmetry is also global and it 
can be understood by splitting $\psi = \psi_R + \psi_L$ into its right and left chiral 
parts,
\begin{equation}
 \psi_R = \frac{1 + \gamma^5}{2}\psi, \quad 
 \psi_L = \frac{1 - \gamma^5}{2}\psi.
 \label{eq:psi5}
\end{equation}
Substituting the above decomposition into Eq.~\eqref{eq:L}, it can 
be seen that 
\begin{equation}
 \cL = \frac{i}{2} (\overline{\psi}_R \overleftrightarrow{\slashed{\partial}} \psi_R + 
 \overline{\psi}_L \overleftrightarrow{\slashed{\partial}} \psi_L),
 \label{eq:L5}
\end{equation}
where $f \overleftrightarrow{\partial_\mu} g = 
f\partial_\mu g - (\partial_\mu f) g$ is the bilateral derivative. 
In analogy to the vector current $J^\mu_V$ introduced in Eq.~\eqref{eq:JV}, 
one can introduce the axial current $J^\mu_A$ via
\begin{equation}
 J^\mu_A = \bar{\psi} \gamma^\mu \gamma^5 \psi,
 \label{eq:JA}
\end{equation}
which corresponds to a conserved axial charge, $\partial_\mu J^\mu_A = 0$, provided $\psi$ satisfies 
the Dirac equation, Eq.~\eqref{eq:diraceq}.

The free theory~\eq{eq:L} is also invariant under a ``helical'' $U(1)_H$ symmetry: 
\begin{align}
U(1)_H:& &
\psi \rightarrow& e^{2 i\alpha_H h} \psi, &
{\bar\psi} \rightarrow& {\bar \psi} e^{- 2 i\alpha_H h},
\label{eq:U1:H}
\end{align}
where the helicity operator,
\beqn
h = \frac{{\bs s}\cdot {\bs p}}{p} 
\equiv
\frac{1}{2p}\gamma^5 H,
\label{eq:h}
\eeqn
is determined via the spin operator $s^i = \frac{1}{2} \varepsilon^{0ijk} \Sigma_{jk}$
which is a part of the covariant spin tensor $\Sigma^{\mu\nu} = \frac{i}{4}[\gamma^\mu,\gamma^\nu]$. The Hamiltonian of free fermions~\eq{eq:L} is:
\beqn
H = -i \gamma^0 {\bs \gamma} \cdot {\bs \nabla} \equiv \gamma^0 {\bs \gamma} \cdot {\bs p}.
\label{eq:H}
\eeqn
In order to understand the $U(1)_H$ transformations in Eq.~\eqref{eq:U1:H}, let us consider a decomposition 
of $\psi = \psi_{\uparrow} + \psi_{\downarrow}$ into its positive and negative helicity components, in 
a manner equivalent to the one in Eq.~\eqref{eq:psi5}:
\begin{equation}
 \psi_{\uparrow} = \left(\frac{1}{2} + h\right) \psi, \quad 
 \psi_{\downarrow} = \left(\frac{1}{2} - h\right) \psi.
 \label{eq:psiH}
\end{equation}
We now seek to obtain a decomposition of the Lagrangian \eqref{eq:L} 
similar to that in Eq.~\eqref{eq:L5},
\begin{equation}
 \cL_H = \frac{i}{2} (\overline{\psi}_\uparrow \overleftrightarrow{\slashed{\partial}} \psi_\uparrow + 
 \overline{\psi}_\downarrow \overleftrightarrow{\slashed{\partial}} \psi_\downarrow).\label{eq:LH}
\end{equation}
As opposed to the chiral case, the above ``helical'' Lagrangian  
differs from the Lagrangian $\cL$ \eqref{eq:L} via
\begin{equation}
 \delta\cL_H = \cL - \cL_H = \frac{i}{2} (\overline{\psi}_\uparrow \overleftrightarrow{\slashed{\partial}} \psi_\downarrow + 
 \overline{\psi}_\downarrow \overleftrightarrow{\slashed{\partial}} \psi_\uparrow). \label{eq:dLH}
\end{equation}
We will now show that the above contribution can be interpreted as 
a 3-divergence which cancels when considering the action $\cS = \int d^4x\,  \cL$. 
For this purpose, it is convenient to introduce the Fourier transform 
of the spinor $\psi$ with respect to the spatial coordinates,
\begin{equation}
 \psi(t, \bm{x}) = \int \frac{d^3p}{(2\pi)^3} e^{i \bm{p} \cdot \bm{x}} 
 \psi_{\bm{p}}(t).
\end{equation}
Using Eq.~\eqref{eq:h}, the positive and negative helicity parts are 
\begin{align}
 \psi_{\uparrow/\downarrow} =& \int \frac{d^3p}{(2\pi)^3} e^{i \bm{p} \cdot \bm{x}} 
 \left(\frac{1}{2} \pm \frac{\bm{s} \cdot \bm{p}}{p} \right) \psi_{\bm{p}}(t).
%
\end{align}
Inserting now the above into \eqref{eq:dLH} and integrating with 
respect to $d^3x$, we arrive at 
\begin{multline}
 \int d^3x\, \delta \cL_H = \sum_{\lambda = \pm 1/2} \int \frac{d^3p}{(2\pi)^3} 
 \bar{\psi}_{\bm{p}} \left(\lambda + \frac{\bm{s} \cdot \bm{p}}{p}\right)\\\times
 \left(\frac{i}{2} \gamma^0 \overleftrightarrow{\partial_t} - \bm{\gamma} \cdot \bm{p}\right)
 \left(\lambda - \frac{\bm{s} \cdot \bm{p}}{p}\right) \psi_{\bm{p}},\label{eq:dLH_aux}
\end{multline}
where the bilateral derivative $\overleftrightarrow{\partial_t}$ acts on the spinors $\psi_{\bm{p}}$ and $\bar{\psi}_{\bm{p}}$.
Writing the spin matrix as
\begin{equation}
 \bm{s} = \frac{1}{2} \gamma^5 \gamma^0 \bm{\gamma},
\end{equation}
it can be seen that 
both $\gamma^0$ and $\bm{\gamma} \cdot \bm{p}$ commute with
with $\bm{s} \cdot \bm{p} / p$. 
Further noting that $[(\bm{s} \cdot \bm{p}) / p]^2 = 1/4$, it 
is straightforward to show that the factor between $\bar{\psi}_{\bm{p}}$ and $\psi_{\bm{p}}$ appearing in Eq.~\eqref{eq:dLH_aux}
vanishes separately for each value of $\lambda$, leading to 
\begin{equation}
 \int d^3x\, \delta \cL_H = 0.
\end{equation}
Coming back to the $U(1)_H$ transformations, these can be 
applied on the spinor $\psi$ at the level of the decomposition 
in Eq.~\eqref{eq:psiH},
\begin{equation}
 e^{2i \alpha_H h} \psi = e^{i \alpha_H} \psi_{\uparrow} + 
 e^{-i \alpha_H} \psi_{\downarrow}.
\end{equation}
Denoting by $\cL' = \cL_H' + \delta \cL_H'$ the form 
of the Lagrangian after the $U(1)_H$ transformation, it can be 
seen that $\cL_H' = \cL_H$, while
\begin{equation}
 \delta \cL_H' = \frac{i}{2}(
 e^{-2i\alpha_H} \bar{\psi}_\uparrow \overleftrightarrow{\slashed{\partial}} \psi_\downarrow + 
 e^{2i\alpha_H} \bar{\psi}_\downarrow \overleftrightarrow{\slashed{\partial}} \psi_\uparrow).
\end{equation}
Under integration with respect to $d^3x$, $\delta \cL_H'$ makes vanishing contributions and thus 
\begin{multline}
 \cS = \int d^4x\, \cL \xrightarrow[]{U(1)_H} \cS' = \int d^4x\, \cL' = \int d^4x\, \cL \\
 = \cS.
 \label{eq:U1H_invariance}
\end{multline}
We are thus lead to interpret the $U(1)_H$ transformations
as non-local. Nevertheless, the symmetry implied by 
Eq.~\eqref{eq:U1H_invariance} allows us to introduce 
the helicity current $J^\mu_H$,
\begin{equation}
 J^\mu_H = \bar{\psi} \gamma^\mu h \psi + (\overline{h \psi}) \gamma^\mu \psi,\label{eq:JH}
\end{equation}
which satisfies $\partial_\mu J^\mu_H = 0$ when $\psi$ is a solution of 
the Dirac equation \eqref{eq:diraceq}. Importantly, we remind here 
that the helicity current $J^\mu_H$ remains conserved when the 
fermions have finite mass $m$, as discussed in Ref.~\cite{Ambrus:2019ayb}. Our paper is devoted to the massless case only.

\subsection{Lorentz covariance of the helicity current $J^\mu_H$} \label{sec:helical:cov}

We now discuss the covariance of the helicity current under Lorentz transformations, using Wigner's method of induced representations, as discussed in Ref.~\cite{Wein95_vol1} (see Chapter 2 therein). As is well-known, the helicity of a massive Dirac particle depends on the reference frame in which it is measured. For example, let us consider a positive-helicity particle travelling along the positive $z$ axis, with four-momentum $p^\mu = (E, 0, 0, p)^\mu$. To an observer also travelling along the positive $z$ axis with velocity $V > p / E$, the particle will appear to move towards the negative $z$ axis. At the same time, in the rest frame of such an observer, the spin of the particle will remain aligned along the positive $z$ axis, since the spin projection along the $z$ axis does not change under boosts along the same axis. Therefore, the observer will measure a negative helicity in their reference frame. 

In contrast, the helicity of a massless particle is Lorentz-invariant. This becomes manifest when considering the transformation properties of the Dirac spinor under a Lorentz transformation $\Lambda$:
\begin{equation}
    \psi(x) \rightarrow \psi^\Lambda(x) = D(\Lambda) \psi(\Lambda^{-1} x),
    \label{eq:psi_Lambda}
\end{equation}
where $D(\Lambda) = e^{-i S_{\alpha\beta} \omega^{\alpha\beta}}$ is the unitary representation of the Lorentz group, while $S^{\alpha\beta} = \frac{i}{4} [\gamma^\alpha,\gamma^\beta]$ are the spin part of the generators of the Lorentz transformations for the Dirac field. For a one-particle momentum eigenstate $U_{p,\lambda}(x)$, satisfying
\begin{align}
 i \partial_\mu U_{p,\lambda}(x) &= U_{p,\lambda}(x) p_\mu, &
 h U_{p,\lambda}(x) &= U_{p,\lambda}(x) \lambda,
\end{align}
one can show that \cite{Wein95_vol1}
\begin{align}
 U_{p,\lambda}(x) \rightarrow U^\Lambda_{p,\lambda}(x) &= D(\Lambda) U_{p,\lambda}(\Lambda^{-1} x) \nonumber\\
 &= \sqrt{\frac{(\Lambda p)^0}{p^0}} e^{i \lambda \theta(\Lambda, p)} U_{\Lambda p, \lambda}(x),
 \label{eq:Uplambda_Lambda}
\end{align}
where $\theta$ is a real number defined by
\beqn 
W(\Lambda, p) & \equiv & L^{-1}(\Lambda p) \Lambda L(p) \nonumber \\
  & \equiv & S[\alpha(\Lambda, p), \beta(\Lambda, p)] R_3[\theta(\Lambda, p)].
 \label{eq:W}
\eeqn
In the above, $W(\Lambda, p)$ represents an element of the little group $ISO(2)$ of the massless Dirac field, while $L(p) = R_3(\varphi_p) R_2(\theta_p) L_3(k \rightarrow p)$ is a pure boost transforming the null vector $k^\mu = (k,0,0,k)^\mu$ into $p^\mu$ characterized by the spherical coordinates $(p, \theta_p, \varphi_p)$.
The transformation $S(\alpha,\beta)$ and the rotation $R_3(\theta)$ are given explicitly by
\begin{align}
 [S(\alpha,\beta)]^\mu{}_\nu &= 
 \begin{pmatrix}
  1 + \zeta & \alpha & \beta & -\zeta \\
  \alpha & 1 & 0 & -\alpha \\ 
  \beta & 0 & 1 & -\beta \\ 
  \zeta & \alpha & \beta & 1 - \zeta
 \end{pmatrix}, \nonumber\\
  [R_3(\theta)]^\mu{}_\nu &= 
 \begin{pmatrix}
   1 & 0 & 0 & 0\\
   0 & \cos \theta & -\sin\theta & 0 \\
   0 & \sin \theta & \cos \theta & 0 \\
   0 & 0 & 0 & 1
 \end{pmatrix},
  \label{eq:Sab}
\end{align}
where $\zeta = \frac{1}{2}(\alpha^2 + \beta^2)$.

Let us now consider the transformation properties of the helicity current. In a reference frame related to the laboratory frame by a Lorentz transformation $\Lambda$, the helicity current is given by
\begin{equation}
 J^{\Lambda; \mu}_H(x) = \overline{\psi^\Lambda(x)} \gamma^\mu h_x \psi^\Lambda(x) + (\overline{h_x \psi^\Lambda(x)}) \gamma^\mu \psi^\Lambda(x).
\end{equation}
To see how $J^{\Lambda;\mu}_H(x)$ is connected to the laboratory-frame helicity current $J^\mu_H(x)$, we first replace $\psi^\Lambda(x)$ as shown in Eq.~\eqref{eq:psi_Lambda}:
\begin{multline}
 J^{\Lambda; \mu}_H(x) = \Lambda^\mu{}_\nu 
 \bar{\psi}(\Lambda^{-1}x) \gamma^\nu \overline{D(\Lambda)} h_x D(\Lambda) \psi(\Lambda^{-1} x)\\
 + \text{h.c.},
 \label{eq:JH_Lambda}
\end{multline}
where $\text{h.c.}$ denotes the Hermitian conjugate and we used the property $\overline{D(\Lambda)} \gamma^\mu D(\Lambda) = \Lambda^\mu{}_\nu \gamma^\nu$.
In order to evaluate $\overline{D(\Lambda)} h_x D(\Lambda) \psi(\Lambda^{-1} x)$, we consider the expansion
\begin{equation}
 \psi(x) = \sum_\lambda \int d^3p [U_{p,\lambda}(x) b_\lambda(p) + V_{p,\lambda}(x) d^\dagger_\lambda(p)],
\end{equation}
where $V_{p,\lambda}(x) = i \gamma^2 U^*_{p,\lambda}(x)$ represent the anti-particle states, satisfying 
\begin{align}
 i \partial_\mu V_{p,\lambda}(x) &= -V_{p,\lambda}(x) p_\mu, &
 h V_{p,\lambda}(x) &= V_{p,\lambda}(x) \lambda.
\end{align}

When applying $\overline{D(\Lambda)} h_x D(\Lambda)$ on $\psi(\Lambda^{-1} x)$, Eq.~\eqref{eq:Uplambda_Lambda} can be employed to replace $D(\Lambda) U_{p,\lambda}(\Lambda^{-1} x)$, leading to
\begin{multline}
 \overline{D(\Lambda)} h_x D(\Lambda) U_{p,\lambda}(\Lambda^{-1} x) \\
 = \lambda \sqrt{\frac{(\Lambda p)^0}{p^0}} e^{i \lambda \theta(\Lambda, p)} \overline{D(\Lambda)} U_{\Lambda p, \lambda}(x),
 \label{eq:Ubar_h_U_Uplambda}
\end{multline}
where we used the property $h_x U_{\Lambda p,\lambda}(x) = U_{\Lambda p, \lambda}(x) \lambda$. Noting that $\overline{D(\Lambda)} = D(\Lambda^{-1})$, Eq.~\eqref{eq:Uplambda_Lambda} implies that
\begin{equation}
 \overline{D(\Lambda)} U_{\Lambda p,\lambda}(x) = 
 \sqrt{\frac{p^0}{(\Lambda p)^0}} e^{i \lambda \theta(\Lambda^{-1}, \Lambda p)} U_{p,\lambda} (\Lambda^{-1} x),
 \label{eq:Ubar_ULambdaplambda}
\end{equation}
where the phase factor $\theta(\Lambda^{-1}, \Lambda p)$ corresponds to the transformation
\begin{equation}
 W(\Lambda^{-1}, \Lambda p) = L^{-1}(p) \Lambda^{-1} L(\Lambda p) = [W(\Lambda, p)]^{-1}.
\end{equation}
The above inverse can be computed as
\begin{equation}
 [W(\Lambda, p)]^{-1} =
 R^{-1}_3[\theta(\Lambda, p)] S^{-1}[\alpha(\Lambda,p),\beta(\Lambda,p)],
\end{equation}
where it is not difficult to see that 
$R^{-1}_3[\theta(\Lambda,p)] = R_3[-\theta(\Lambda,p)]$ and $S^{-1}[\alpha(\Lambda,p),\beta(\Lambda,p)] = S[-\alpha(\Lambda,p),-\beta(\Lambda,p)]$. Using now the property
\begin{multline}
 R^{-1}_3(\theta) S(\alpha,\beta) R_3(\theta) \\
 = S(\alpha \cos\theta + \beta \sin\theta, \beta \cos\theta - \alpha \sin\theta),
\end{multline}
the matrix $W(\Lambda^{-1},\Lambda p)$ can be put in the form \eqref{eq:W} with parameters
\begin{equation}
 \begin{pmatrix}
  \alpha(\Lambda^{-1}, \Lambda p) \\
  \beta(\Lambda^{-1}, \Lambda p)
 \end{pmatrix} = 
 \begin{pmatrix}
   \cos\theta & \sin \theta \\
   -\sin\theta & \cos\theta
 \end{pmatrix}
 \begin{pmatrix}
  -\alpha(\Lambda,p)\\
  -\beta(\Lambda,p)
 \end{pmatrix},
\end{equation}
while $\theta(\Lambda^{-1}, \Lambda p) = -\theta(\Lambda,p)$. Using this latter result together with Eq.~\eqref{eq:Ubar_ULambdaplambda} into Eq.~\eqref{eq:Ubar_h_U_Uplambda}, we arrive at
\begin{equation}
 \overline{D(\Lambda)} h_x D(\Lambda) U_{p,\lambda}(\Lambda^{-1} x) = U_{p, \lambda}(\Lambda^{-1} x) \lambda.
\end{equation}
Noting that a similar relation holds also for the antiparticle states $V_{p,\lambda}(\Lambda^{-1} x)$, we find
\begin{multline}
 \overline{D(\Lambda)} h_x D(\Lambda) \psi(\Lambda^{-1} x) \\
 = \sum_\lambda \lambda \int d^3p [U_{p,\lambda}(\Lambda^{-1} x) b_\lambda(p) + V_{p,\lambda}(\Lambda^{-1} x) d^\dagger_\lambda(p)]\\
 = h_{\Lambda^{-1} x} \psi(\Lambda^{-1} x).
\end{multline}
With the above relation, we can conclude that for massless particles, $J^\mu_H(x)$ transforms covariantly as a four-vector under the Lorentz group transformations:
\begin{equation}
 J^\mu_H(x) \rightarrow J^{\Lambda,\mu}_H(x) = \Lambda^\mu{}_\nu J^\nu_H(\Lambda^{-1} x).
\end{equation}

\subsection{Second quantisation and the $\widehat{Q}_{V/A/H}$ 
charge operators} \label{sec:helical:Q}

Since $\partial_\mu J^\mu_{V/A/H}= 0$ for the classical solutions of the free-fermion theory~\eq{eq:L}, the vector-axial-helical charges,
\begin{equation}
 \begin{pmatrix}
  Q_V \\ Q_A \\ Q_H
 \end{pmatrix} = 
 \int d^3x \,
 {\bar \psi} \gamma^0
 \begin{pmatrix}
  1 \\ \gamma^5 \\ 2h
 \end{pmatrix} \psi,
 \label{eq:charges}
\end{equation}
form a ``triad'' of the classically conserved $U(1)$ quantities of the massless Dirac fermions~\cite{footnote:2}.

The Dirac equation \eqref{eq:diraceq} admits a complete set of solutions comprised of the particle and antiparticle modes, $U_j$ and $V_j = i \gamma^2 U_j^*$, where $j$ cumulatively indexes the eigenmodes. The commutation relations $[\gamma^5, h] = [H, \gamma^5] = [H, h] = 0$ indicate that these modes are simultaneous eigenfunctions of the Hamiltonian, chirality and helicity operators:
\beqs
\begin{align}
H U_j = & E_j U_j, & 
H V_j =& -E_j V_j,\\
\gamma^5 U_j =& \chi_j U_j, &
\gamma^5 V_j =& -\chi_j V_j,\\
 h U_j =& \lambda_j U_j, &
 h V_j =& \lambda_j V_j,
\end{align}
\label{eq:UV:chi:h}
\eeqs
where $E_j = p_j$ is the (positive) mode energy. The relation between the chirality $\chi_j = \pm1$ and helicity $\lambda_j = \pm\frac{1}{2}$ can be established using Eq.~\eqref{eq:h}:
\begin{equation}
 \chi_j = 2\lambda_j.
 \label{eq:lambda_chi}
\end{equation}
The field operator is 
\begin{equation}
 \widehat{\psi}(x) = \sum_j [U_j(x) \hat{b}_j + V_j(x) \hat{d}^\dagger_j],
 \label{eq:psi}
\end{equation}
where canonical anticommutation rules for the particle $\hat{b}_j$ and antiparticle $\hat{d}_j$ operators are assumed (we use hats to denote operators acting on Fock space). Since we consider $E_j > 0$, we are essentially working with the maximally-symmetric (Minkowski) vacuum, corresponding to an inertial Lorentz frame. Other choices are possible, for example the
rotating vacuum introduced by Iyer \cite{iyer82}, in which the modes with positive co-rotating energy $\widetilde{E}_j = E_j - \Omega m_j$ correspond to particle modes ($m_j$ is the total angular momentum along the axis rotation). The rotation parameter $\Omega$ will be introduced below. 

The operators of the conserved charges~\eqref{eq:charges} are:
\beqs
\beqn
 :\widehat{Q}_V: &=& \sum_j (\hat{b}^\dagger_j \hat{b}_j - 
 \hat{d}^\dagger_j \hat{d}_j),
 \label{eq:Q:V}\\
 :\widehat{Q}_A: &=& \sum_j \chi_j (\hat{b}^\dagger_j \hat{b}_j + \hat{d}^\dagger_j \hat{d}_j),
 \label{eq:Q:A}\\
 :\widehat{Q}_H: &=& \sum_j 2\lambda_j (\hat{b}^\dagger_j \hat{b}_j - \hat{d}^\dagger_j \hat{d}_j),
 \label{eq:Q:H}
\eeqn
 \label{eq:Q}
\eeqs
where the colons denote Wick (normal) ordering. The particle and antiparticle states contribute differently to the axial~\eq{eq:Q:A} and helicity~\eq{eq:Q:H} charges. By virtue of Eq.~\eqref{eq:lambda_chi}, Eqs.~\eq{eq:UV:chi:h} and \eq{eq:Q} also imply that the chiralities and helicities are indeed equal (opposite) to each other for particle (antiparticle) modes.

\subsection{Discrete symmetries of the 
$\widehat{J}^\mu_{V/A/H}$ charge currents}\label{sec:helical:CPT}

In this section, the transformation properties of the helicity charge current
(HCC) operator under the charge conjugation $\mathcal{C}$, parity $\mathcal{P}$, and time reversal
$\mathcal{T}$ operations are derived, following the conventions of Ref.~\cite{itzykson80}.
In order to introduce the helicity charge current operator, $\widehat{J}^\mu_H(x)$, 
we first note that the classical quantity $2h \psi(x)$ can be decomposed as
\begin{equation}
 2 h \psi(x) = \sum_j 2\lambda_j (U_j b_j + V_j d_j^*).
\end{equation}
After second quantisation, the equivalent relation 
can be obtained by taking the commutator of the quantum operator $\widehat{\psi}$ with the helicity charge operator, 
$\widehat{Q}_H$, defined in Eq.~\eqref{eq:Q:H}:
\begin{equation}
 [\widehat{\psi}(x), \widehat{Q}_H] = \sum_j 2\lambda_j 
 (U_j \hat{b}_j + V_j \hat{d}_j^\dagger).
\end{equation}
Based on the classical definition \eqref{eq:JH} of the helicity charge operator, 
$J^\mu_H$, its quantum form can be introduced as follows:
    \begin{equation}
 \widehat{J}^\mu_H = \frac{1}{4}\left([[\widehat{Q}_H, \widehat{\overline{\psi}}], 
 \gamma^\mu \widehat{\psi}] - 
 [\widehat{\overline{\psi}}, [\widehat{Q}_H, \gamma^\mu \widehat{\psi}]]\right).
\end{equation}
The above definition is exactly mirrored in the case of the vector charge current:
\begin{equation}
 \widehat{J}^\mu_V = \frac{1}{4}\left([[\widehat{Q}_V, \widehat{\overline{\psi}}], 
 \gamma^\mu \widehat{\psi}] - 
 [\widehat{\overline{\psi}}, [\widehat{Q}_V, \gamma^\mu \widehat{\psi}]]\right),
\end{equation}
where the commutators of the field operator and its adjoint with the vector charge 
operator are trivially
\begin{equation}
 [\widehat{\psi}, \widehat{Q}_V] = \widehat{\psi}, \qquad 
 [\widehat{\overline{\psi}}, \widehat{Q}_V] = -\widehat{\overline{\psi}}.
\end{equation}

The CPT symmetries of the HCC can be described in close analogy 
to those of the VCC. Specifically, we are interested in computing the 
following quantities:
\begin{equation}
 \widehat{\mathcal{C}}\, \widehat{J}^\mu_H(x) \widehat{\mathcal{C}}^\dagger, \qquad 
 \widehat{\mathcal{P}} \widehat{J}^\mu_H(x) \widehat{\mathcal{P}}^\dagger, \qquad
 \widehat{\mathcal{T}} \widehat{J}^\mu_H(x) \widehat{\mathcal{T}}^\dagger,
\end{equation}
where the charge conjugation ($\widehat{\mathcal{C}}$), parity ($\widehat{\mathcal{P}}$) and 
time reversal ($\widehat{\mathcal{T}}$) operators are assumed to be unitary.
Furthermore, in the Dirac theory, these operators have the following action 
on the one-particle operators \cite{itzykson80}:
\begin{align}
 \widehat{\mathcal{C}} \hat{b}_j \widehat{\mathcal{C}}^\dagger =& \eta_C \hat{d}_j, &
 \widehat{\mathcal{C}} \hat{d}_j \widehat{\mathcal{C}}^\dagger =& \eta_C^* \hat{b}_j, \nonumber\\
 \widehat{\mathcal{P}} \hat{b}_j \widehat{\mathcal{P}}^\dagger =& \eta_P \hat{b}_{j_P},&
 \widehat{\mathcal{P}} \hat{d}_j \widehat{\mathcal{P}}^\dagger =& -\eta_P^* \hat{d}_{j_P},\\
 \widehat{\mathcal{T}} \hat{b}_j \widehat{\mathcal{T}}^\dagger =& \eta_T \hat{b}_{j_T} e^{-i \theta_{b,j}}, &
 \widehat{\mathcal{T}} \hat{d}_j \widehat{\mathcal{T}}^\dagger =& \eta_T^* \hat{d}_{j_T} e^{i \theta_{d,j}}, \nonumber
\end{align}
where 
$\eta_C$, $\eta_P$, $\eta_T$, $e^{-i \theta_{b,j}}$ and 
$e^{i \theta_{d,j}}$ are irrelevant phases.
For the set of eigenvalues $j \equiv (E_j, k_j, m_j, \lambda_j)$, 
the parity transformation gives $j_P \equiv (E_j, -k_j, m_j, -\lambda_j)$, while 
the time reversal gives $j_T \equiv (E_j, -k_j, -m_j, \lambda_j)$.

\begin{table}[t]
\begin{center}
\begin{tabular}{|c||c|c|c||c|c|c|}
\hline
 	&   $Q_V$ & 	$Q_A$ 	& 	$Q_H$	&   ${\bs J}_V$ 	&	${\bs J}_A$ 
 	& ${\bs J}_H$ \\
\hline
$C$	&   $-$	&	$+$	&	$-$ &	$-$	&	$+$	&	$-$	
\\
\hline
$P$	&   $+$	&	$-$	&	$-$ &	$-$	&	$+$	&	$+$
\\
\hline
$T$	&   $+$	&	$+$	&	$+$ &	$-$	&	$-$	&	$-$	
\\
\hline
\end{tabular}
\end{center}
\vskip 3mm
\caption{
Behaviour of the vector $(V)$, axial~$(A)$, and helical $(H)$ charges ($Q$) and currents (${\bs J}$) of a massless Dirac fermion under the $C$-, $P$-, and $T$-inversions. The signs $+/-$ indicate the even/odd nature of these quantities under the corresponding discrete transformations.
}
\label{tbl:inversions}
\end{table}

Let us now consider the action of a discrete symmetry, $\widehat{\mathcal{S}}$, on the 
helicity and vector charge currents:
\begin{multline}
 \widehat{\mathcal{S}} \,\widehat{J}^\mu_{V/H} \widehat{\mathcal{S}}^\dagger
 = \frac{1}{4} \widehat{\mathcal{S}} \left([[\widehat{Q}_{V/H}, \widehat{\overline{\psi}}], 
 \gamma^\mu \widehat{\psi}]\right.\\
 \left. - 
 [\widehat{\overline{\psi}}, [\widehat{Q}_{V/H}, \gamma^\mu \widehat{\psi}]]\right)
 \widehat{\mathcal{S}}^\dagger.\label{eq:SJVHS}
\end{multline}
The symmetry transformation can be applied to the commutator of two operators, 
$\widehat{A}$ and $\widehat{B}$, using the following rule:
\begin{equation}
 \widehat{\mathcal{S}} [\widehat{A}, \widehat{B}] \widehat{\mathcal{S}}^\dagger = 
 [\widehat{\mathcal{S}} \,\widehat{A}\, \widehat{\mathcal{S}}^\dagger, 
 \widehat{\mathcal{S}} \,\widehat{B}\, \widehat{\mathcal{S}}^\dagger].
\end{equation}
Looking again at Eq.~\eqref{eq:SJVHS}, it can be seen that the CPT properties 
of the HCC can be inferred from those of the VCC if the transformation properties 
of the corresponding charge operators, $\widehat{Q}_V$ and $\widehat{Q}_H$, are 
known. 
It is not difficult to show that
\begin{align}
 \widehat{\mathcal{C}}\, \widehat{Q}_V \widehat{\mathcal{C}}^\dagger =& - \widehat{Q}_V, &
 \widehat{\mathcal{C}}\, \widehat{Q}_H \widehat{\mathcal{C}}^\dagger =& -\widehat{Q}_H, \nonumber\\
 \widehat{\mathcal{P}}\, \widehat{Q}_V \widehat{\mathcal{P}}^\dagger =& \widehat{Q}_V, &
 \widehat{\mathcal{P}}\, \widehat{Q}_H \widehat{\mathcal{P}}^\dagger =& -\widehat{Q}_H, \nonumber\\
 \widehat{\mathcal{T}}\, \widehat{Q}_V \widehat{\mathcal{T}}^\dagger =& \widehat{Q}_V, &
 \widehat{\mathcal{T}}\, \widehat{Q}_H \widehat{\mathcal{T}}^\dagger =& \widehat{Q}_H.
 \label{eq:QH_CPT}
\end{align}
The properties of $\widehat{Q}_H$ differ from those of $\widehat{Q}_V$ due 
to the behaviour of the helicity, $\lambda_j$, under the CPT transformations,
as can be seen in the left half of Table~\ref{tbl:inversions}.

Using Eq.~\eqref{eq:QH_CPT}, the CPT properties of the HCC can 
be inferred from those of the VCC, as follows:
\begin{align}
 \widehat{\mathcal{C}}\, \widehat{J}^\mu_V(x) \widehat{\mathcal{C}}^\dagger =& -\widehat{J}^\mu_V(x), &
 \widehat{\mathcal{C}}\, \widehat{J}^\mu_H(x) \widehat{\mathcal{C}}^\dagger =& -\widehat{J}^\mu_H(x), \nonumber\\
 \widehat{\mathcal{P}}\, \widehat{J}^\mu_V(x) \widehat{\mathcal{P}}^\dagger =& \widehat{J}^{\,V}_\mu(\widetilde{x}), &
 \widehat{\mathcal{P}}\, \widehat{J}^\mu_H(x) \widehat{\mathcal{P}}^\dagger =& -\widehat{J}^{\,H}_\mu(\widetilde{x}), \nonumber\\
 \widehat{\mathcal{T}}\, \widehat{J}^\mu_V(x) \widehat{\mathcal{T}}^\dagger =& \widehat{J}^{\,V}_\mu(-\widetilde{x}), &
 \widehat{\mathcal{T}}\, \widehat{J}^\mu_H(x) \widehat{\mathcal{T}}^\dagger =& \widehat{J}^{\,H}_\mu(-\widetilde{x}), 
\end{align}
where $\widetilde{x} = (t, -\bm{x})$ when $x = (t, \bm{x})$.
When $\mu = i$ is a spatial index, $\widehat{J}^H_i = -\widehat{J}^i_H$, so that 
the temporal and spatial parts of the HCC are even under the $T$ and $P$ transformations, respectively, as 
can be seen in the right half of Table~\ref{tbl:inversions}.

\subsection{Helical chemical potential} \label{sec:helical:muH}


The chirality and helicity are different quantities. 
The chirality of a particle is equal to its helicity (a right-chiral particle has a right-handed helicity) while the chirality of an antiparticle is opposite to its helicity (a right-chiral antiparticle has a left-handed helicity)~\cite{peskin95}. For a single fermion the vector (particle/antiparticle) and the axial (right-/left-chiral) charges determine uniquely its helicity. This is no longer the case for a fermion ensemble.



In a generic ensemble of Dirac fermions, all three classically conserved charges, $Q_\ell = \langle {\widehat{Q}_\ell}\rangle$, can be expressed via linear combinations of four independent numbers that count particles ($N^R_\uparrow$ and $N^L_\downarrow$) and antiparticles (${\bar N}^R_\downarrow$ and ${\bar N}^L_\uparrow$). Let us attribute, for a moment, a ``chemical potential'' to each of the mentioned individual numbers ($\mu^R_\uparrow$ is associated with $N^R_\uparrow$, etc). The system can be described both via the $V$, $A$, $H$ charges and alternatively, via the particle numbers:
\begin{align}
\delta \cL_Q &= \mu_V Q_V  + \mu_A Q_A + \mu_H Q_H \nonumber \\
&= \mu^{R}_\uparrow N^{R}_\uparrow  + \mu^{L}_\downarrow N^{L}_\downarrow 
        + {\bar \mu}^{R}_\downarrow {\bar N}{}^{R}_\downarrow  + {\bar \mu}^{L}_\uparrow {\bar N}{}^{L}_\uparrow.
\label{eq:delta:L}
\end{align}
The individual particles contribute to each charge differently: 
\begin{subequations}
\begin{align}
 Q_V =& (N^R_\uparrow + N^L_\downarrow) - ({\bar N}{}^R_\downarrow + {\bar N}{}^L_\uparrow), \\
 Q_A =& (N^R_\uparrow + {\bar N}{}^L_\uparrow) - ( N^L_\downarrow + {\bar N}{}^R_\downarrow), \\
 Q_H =& (N^R_\uparrow + {\bar N}{}^R_\downarrow) - ( N^L_\downarrow + {\bar N}{}^L_\uparrow).
\end{align}
\end{subequations}
Then Eq.~\eq{eq:delta:L} gives us the relations:
\begin{subequations}
\begin{align}
\mu_V =& \left[ \left( \mu^R_\uparrow + \mu^L_\downarrow \right) - \left( {\bar \mu}^R_\downarrow + {\bar \mu}^L_\uparrow \right) \right]/4,
\label{eq:mu:V}\\
\mu_A =&
\left[ \left( \mu^R_\uparrow + {\bar \mu}^L_\uparrow \right) - \left( \mu^L_\downarrow + {\bar \mu}^R_\downarrow\right) \right]/4,
\label{eq:mu:A} \\
\mu_H =& 
\left[ \left( \mu^R_\uparrow + {\bar \mu}^R_\downarrow \right) - \left( \mu^L_\downarrow + {\bar \mu}^L_\uparrow \right) \right]/4,
\label{eq:mu:H}
\end{align}
\end{subequations}
that are readily understood.

\begin{figure}[t]
\begin{center}
\begin{tabular}{c}
\includegraphics[width=0.9\columnwidth]{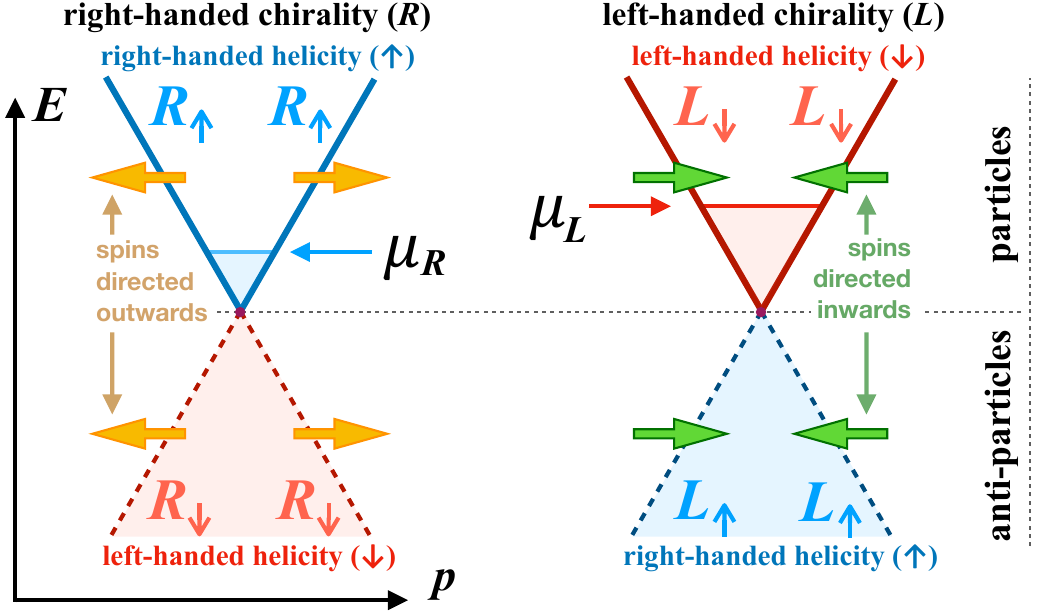} \\
(a) \\
\includegraphics[width=0.9\columnwidth]{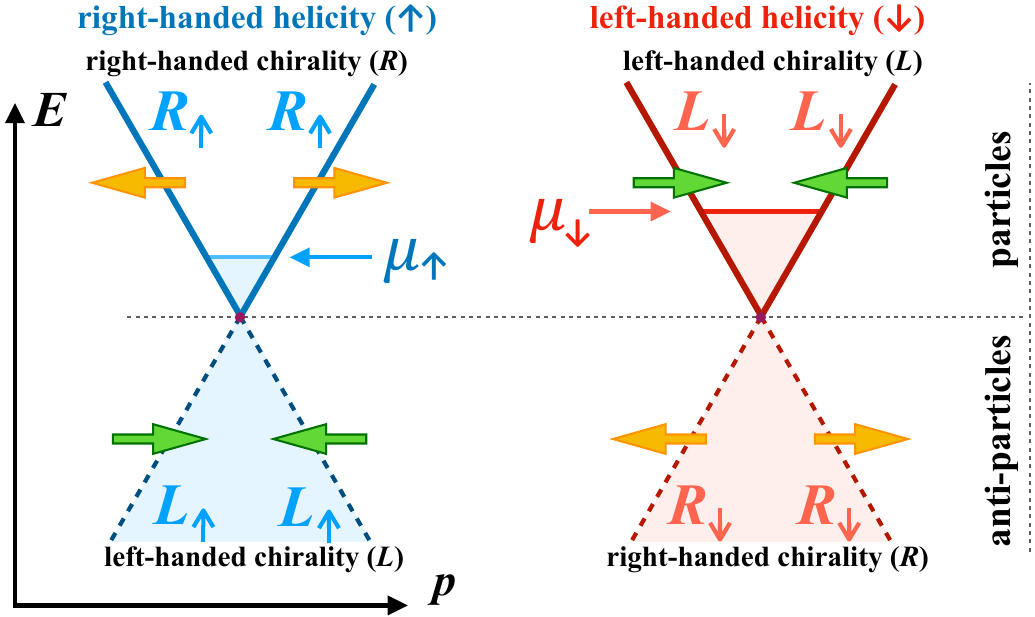}  \\
(b)
\end{tabular}
\end{center}
\vskip 2mm
\caption{The dispersion relations $E_{\bs p} = \pm \lvert{\bs p}\rvert$ for Dirac fermions in (a) chiral and (b) helical basis. The chemical potentials $\mu_{R/L}$ determine the occupation numbers for the right-/left-handed chiralities, while the chemical potentials $\mu_{\uparrow/\downarrow}$ dictate the densities of the right-/left-handed helicities. The spin orientations $\bs s$ with respect to the fermion's momentum $\bs p$ are shown by the horizontal arrows for each energy branch.}
\label{fig:cones}
\end{figure}

Usually, we focus on chirality and neglect helicity. Then the right- and left-handed chiral chemical potentials are given by the sum of particle and antiparticle contributions~\eq{eq:mu:A}, $\mu_R = (\mu^R_\uparrow - {\bar \mu}^R_\downarrow)/2$ and $\mu_L = (\mu^L_\downarrow - {\bar \mu}^L_\uparrow)/2$, shown in the chiral-cone representation of the energy dispersion in Fig.~\ref{fig:cones}(a). The axial chemical potential takes the familiar form: $\mu_A = (\mu_R - \mu_L)/2$.

If we care about the helicity while being uninterested
in the chirality, 
Eq.~\eq{eq:mu:H} gives the right- and left-handed helical chemical potentials, respectively: $\mu_\uparrow = (\mu^R_\uparrow - {\bar \mu}^L_\uparrow)/2$ and $\mu_\downarrow = (\mu^L_\downarrow - {\bar \mu}^R_\downarrow)/2$. These potentials can be used to label the occupation numbers of the ``helical Weyl cones'', thus characterizing the energy branches in a chirality-independent (and non-equivalent) manner, as shown in Fig.~\ref{fig:cones}(b).  The helical chemical potential takes the suggestive form: $\mu_H = (\mu_\uparrow - \mu_\downarrow)/2$.

Physically, the property of helicity stands in between the vector and the axial characteristics of an ensemble of Dirac fermions: the helical charge (current) density  is a difference between the axial charge (current) density of positive and negative energy eigenstates in the ensemble.

In practice, the similarities and differences between helicity and chirality~\cite{peskin95} imply that an ensemble which contains particles only should possess a total chirality which exactly equals the total helicity of this particle-only ensemble. However, an ensemble of particles and antiparticles does not have the same chirality and helicity. Therefore, as both quantities are conserved in the massless theory, one needs to introduce separate chemical potentials for helicity and chirality to describe the grand canonical ensemble for this system.

\section{Relativistic kinetic theory model}\label{sec:RKT}

In this section, we consider an analysis of fluids possessing
vector, axial and helical charge imbalance under the assumption
of global thermodynamic equilibrium. In Subsec.~\ref{sec:RKT:VAH}, 
we introduce Fermi-Dirac distribution functions to model the 
four types of populations, corresponding to particles and 
antiparticles of positive and negative helicity, serving as 
starting point for the computation of the pressure $P$. We further introduce 
derivatives of the pressure of the form 
$\partial^{n+1} P / \partial^n \mu_A \partial \mu_\ell$, where the case 
$n =0$ corresponds to the familiar charge densities $Q_\ell$, while at 
$n > 0$ we uncover the vortical, circular and shear conductivities 
$\sigma^\omega_\ell$, $\sigma^\tau_\ell$ and $\sigma^\Pi_\ell$
appearing in anomalous transport.
 In Subsec.~\ref{sec:RKT:rot},
we discuss the kinematics of rigid motion, which will prove useful 
when discussing the vortical effects and corrections revealed by means
of a full quantum field theoretical treatment in Sec.~\ref{sec:QFT}.

\subsection{Classical V/A/H fluids}\label{sec:RKT:VAH}

Let us consider an effective kinetic theory description of a fluid
comprised of spin $1/2$ particles and their corresponding antiparticles,
described by a set of four distribution functions, 
\begin{equation}
 f^{{\rm eq};\sigma}_{\bp,\lambda} = \left[ 
 \exp\left(\frac{p \cdot u - \bm{q}_{\sigma,\lambda} \cdot \bm{\mu}}{T}\right) 
 + 1 \right]^{-1},\label{eq:feq}
\end{equation}
where $p$ is the particle four-momentum ($p^2 = 0$ for massless constituents), 
$\sigma = \pm 1$ distinguishes between particles ($+1$) and antipartices ($-1$), 
$\lambda = \pm1/2$ denotes the polarization and $u^\mu$ is the local velocity 
of the fluid. 
The $V/A/H$ chemical potentials appearing in Eq.~\eqref{eq:feq}
are contained in the three-vector $\bm{\mu} = (\mu_V, \mu_A, \mu_H)$ and 
$\bm{q}_{\sigma,\lambda}$ represent the eigenvalues of the $:\widehat{Q}_{V/A/H}:$
charge operators, namely
\begin{align}
 q^V_{\sigma,\lambda} =& \sigma, &
 q^A_{\sigma,\lambda} =& 2\lambda, &
 q^H_{\sigma,\lambda} =& 2\sigma \lambda.
 \label{eq:q_def}
\end{align}
Under the assumption of global thermodynamic equilibrium, the local temperature and chemical 
potentials are related to the macroscopic fluid motion via
\begin{equation}
 T = \Gamma T_0, \qquad \mu_\ell = \Gamma \mu_{\ell;0},
 \label{eq:Tolman}
\end{equation}
where $\Gamma = u^0$ is the local Lorentz factor of the fluid, such that 
the constants $T_0$ and $\mu_{\ell;0}$ 
represent the values of the temperature and chemical potentials where 
the fluid is at rest.

Using the distribution in Eq.~\eqref{eq:feq}, the stress-energy tensor (SET) 
and charge currents can be computed via
\begin{equation}
 \begin{pmatrix}
  T_{\rm cl}^{\mu\nu} \\ J_{\ell, \mathrm{cl}}^\mu 
 \end{pmatrix} = \sum_{\sigma = \pm 1} \sum_{\lambda = \pm \frac{1}{2}} 
 \int dP
 \begin{pmatrix}
  p^\mu p^\nu \\ 
  q^\ell_{\sigma,\lambda} p^\mu
 \end{pmatrix} f^{{\rm eq};\sigma}_{\bp,\lambda},
\end{equation}
where $dP = d^3p / [p^0 (2\pi)^3]$ and the subscript 
``cl''
indicates classical quantities.
Due to the form of $f^{eq}_{\sigma,\lambda}$, it is easy to see that 
both $T_{\rm cl}^{\mu\nu}$ and $J_{\rm cl}^\mu$ take the perfect fluid form,
\begin{equation}
 T_{\rm cl}^{\mu\nu} = (E_{\rm cl} + P_{\rm cl}) u^\mu u^\nu - P_{\rm cl} g^{\mu\nu}, \quad 
 J_{\ell; {\rm cl}}^\mu = Q_{\ell;{\rm cl}} u^\mu,\label{eq:RKT_pf}
\end{equation}
where the energy density for massless particles satisfies $E_{\rm cl} = 3P_{\rm cl}$, while 
the pressure $P_{\rm cl}$ and charge densities $Q_{\ell; {\rm cl}}$ are
\begin{align}
 \begin{pmatrix}
  P_{\rm cl} \\ Q_{\ell; {\rm cl}}
 \end{pmatrix} = \sum_{\sigma,\lambda} \int dP 
 \begin{pmatrix}
  (p \cdot u)^2 / 3\\ 
  q^\ell_{\sigma,\lambda} (p \cdot u)
 \end{pmatrix} f^{{\rm eq};\sigma}_{\bp,\lambda}.
 \label{eq:RKT_P_Q}
\end{align}
Taking into account the expression \eqref{eq:feq} of $f^{{\rm eq};\sigma}_{\bp,\lambda}$, it 
can be seen that the charge densities can be obtained from the pressure via 
\begin{equation}
 Q_{\ell; {\rm cl}} = \frac{\partial P_{\rm cl}}{\partial \mu_\ell}.
\end{equation}
As we will show later in this paper, the higher order derivatives of $P$ 
with respect to the axial chemical potential $\mu_A$ can be related to the
conductivities arising in anomalous transport. In particular, 
we identify the vortical and circular conductivities,
$\sigma^\omega_{\ell;{\rm cl}}$ and $\sigma^\tau_{\ell;{\rm cl}}$, 
as well as the shear conductivities $\sigma^\Pi_{\ell;{\rm cl}}$ as follows:
\begin{gather}
 \sigma^\omega_{\ell;{\rm cl}} = \frac{1}{2} 
 \frac{\partial^2 P_{\rm cl}}{\partial \mu_A \partial \mu_\ell}, \quad
 \sigma^\tau_{\ell;{\rm cl}} = \frac{1}{12} 
 \frac{\partial^3 P_{\rm cl}}{\partial^2 \mu_A \partial \mu_\ell}, \nonumber\\
 \sigma^\Pi_{\ell;{\rm cl}} = \frac{1}{2}
 \frac{\partial^4 P_{\rm cl}}{\partial^3 \mu_A \partial \mu_\ell}.
 \label{eq:RKT_sigmas_P}
\end{gather}

We now compute the pressure by noting that
since both chirality and helicity are frame-independent properties 
for massless particles, the Lorentz boost to the local rest frame can 
be performed as usual:
\begin{equation}
 P_{\rm cl} =  \sum_{\sigma,\lambda} \int_0^\infty \frac{dp\, p^3}{6\pi^2}
 \left[ \exp\left(\frac{p - \bm{q}_{\sigma,\lambda} \cdot \bm{\mu}}{T}\right) 
 + 1 \right]^{-1}.
 \label{eq:RKT_P_aux}
\end{equation}
The integration with respect to $p$ can be performed in terms of the 
polylogarithm function ${\rm Li}_s(z) = \sum_{n = 1}^\infty z^n / n^s$ \cite{olver10} and the results for 
$P_{\rm cl}$, $Q_{\ell; {\rm cl}}$, $\sigma^\omega_{\ell;{\rm cl}}$,
$\sigma^\tau_{\ell;{\rm cl}}$ and $\Pi_{\ell;{\rm cl}}$ 
can be written compactly as
\begin{align}
 P_{\rm cl} =& -\frac{T^4}{\pi^2} \sum_{\sigma,\lambda} 
 {\rm Li}_4(-e^{\alpha_{\sigma,\lambda}}), \nonumber\\
 Q_{\ell; {\rm cl}} =& -\frac{T^3}{\pi^2} \sum_{\sigma,\lambda}
 q^\ell_{\sigma,\lambda} {\rm Li}_3(-e^{\alpha_{\sigma,\lambda}}), \nonumber\\
 \sigma^\omega_{\ell;{\rm cl}} =& -\frac{T^2}{2\pi^2} \sum_{\sigma,\lambda} 
 q^\ell_{\sigma,\lambda} q^A_{\sigma,\lambda} {\rm Li}_2(-e^{\alpha_{\sigma,\lambda} }),\nonumber\\
 \sigma^\tau_{\ell;{\rm cl}} =& 
 -\frac{T}{12\pi^2} \sum_{\sigma,\lambda} q^\ell_{\sigma,\lambda} 
 {\rm Li}_1(-e^{\alpha_{\sigma,\lambda}}),\nonumber\\
 \sigma^{\Pi}_{\ell;{\rm cl}} =& -\frac{1}{2\pi^2} \sum_{\sigma,\lambda} 
 q^\ell_{\sigma,\lambda} q^A_{\sigma,\lambda}
 {\rm Li}_0(-e^{\alpha_{\sigma,\lambda}}),\label{eq:RKT_Li}
\end{align}
where ${\rm Li}_1(-e^\alpha) = -\ln(1 + e^\alpha)$ and 
${\rm Li}_0(-e^\alpha) = -(1 + e^{-\alpha})^{-1}$.
The notation $\alpha_{\sigma,\lambda} = \bm{q}_{\sigma,\lambda} \cdot \bm{\mu} / T$
was introduced for brevity.
In the above expressions, we have affixed the label 
``${\rm cl}$'' to quantities which are inherently quantum,
in order to distinguish the above ``thermodynamic'' 
contributions from quantum corrections which will be 
discussed in the following section (see also the Appendix).
It is worthwhile remarking that the 
computation of $\sigma^\omega_{\ell;{\rm cl}}$ and $\sigma^\Pi_{\ell;{\rm cl}}$ 
involves the product
$q^A_{\sigma,\lambda} q^\ell_{\sigma,\lambda}$, which is equal to $1$
when $\ell = A$. When $\ell \in \{V,H\}$, this product becomes
\begin{align}
 q^A_{\sigma,\lambda} q^V_{\sigma,\lambda} &= 2 \sigma \lambda = q^H_{\sigma,\lambda}, \nonumber\\
 q^A_{\sigma,\lambda} q^H_{\sigma,\lambda} &= 2 \lambda \times 2 \sigma \lambda = q^V_{\sigma,\lambda}.\label{eq:qAql}
\end{align}

In the large temperature limit when $\mu_\ell / T \ll 1$, 
the pressure can be expanded as
\begin{subequations}\label{eq:RKT_highT}
\begin{multline}
 P_{\rm cl} = \frac{7\pi^2 T^4}{180} + 
 \frac{T^2 \bm{\mu}^2}{6} + 
 \frac{4 \mu_V \mu_A \mu_H T}{\pi^2} \ln 2 \\
 + \frac{\mu_V^4 + \mu_A^4 + \mu_H^4 + 
 6(\mu_V^2 \mu_A^2 + \mu_V^2 \mu_H^2 + \mu_A^2\mu_H^2)}{12\pi^2}\\
 + O\left(\frac{\mu_V\mu_A \mu_H \bm{\mu}^2}{T}\right),
 \label{eq:RKT_P}
\end{multline}
where $\bm{\mu}^2 = \mu_V^2 + \mu_A^2 + \mu_H^2$.
The $O(T^{-1})$ terms vanish when either one of $\mu_V$, $\mu_A$ and $\mu_H$ vanishes. The charge densities can be written explicitly as
\begin{align}
 Q_{V; {\rm cl}} \simeq& \frac{\mu_V T^2}{3} + \frac{4\mu_A \mu_H T}{\pi^2} \ln 2 \nonumber \\
 & + 
 \frac{\mu_V(\mu_V^2 + 3 \mu_A^2 + 3 \mu_H^2)}{3\pi^2} 
 ,\nonumber\\
 Q_{A; {\rm cl}} \simeq& \frac{\mu_A T^2}{3} + \frac{4\mu_H \mu_V T}{\pi^2} \ln 2 \nonumber \\
  & + 
 \frac{\mu_A(\mu_A^2 + 3 \mu_H^2 + 3 \mu_V^2)}{3\pi^2} 
 ,\nonumber\\
 Q_{H; {\rm cl}} \simeq& \frac{\mu_H T^2}{3} + \frac{4\mu_V \mu_A T}{\pi^2} \ln 2 \nonumber \\
 & + 
 \frac{\mu_H(\mu_H^2 + 3 \mu_V^2 + 3 \mu_A^2)}{3\pi^2} 
 .
 \label{eq:RKT_Ql}
\end{align}
The above relations neglect $O(T^{-1})$ contributions to $Q_{\ell;{\rm cl}}$ which are proportional to 
$\mu_V \mu_A \mu_H / \mu_\ell$. 
For example, $Q_{V;{\rm cl}}$ receives
the extra corrections $\mu_A \mu_H(\mu_A^2 + \mu_H^2) / 6\pi^2 T + O(T^{-3})$.
The vortical conductivities can be expanded at high temperatures as
\begin{align}
 \sigma^\omega_{V;{\rm cl}} =& 
 \frac{2\mu_H T}{\pi^2} \ln 2 + \frac{\mu_V \mu_A}{\pi^2} +
 O\left(\frac{\mu_H \bm{\mu}^2}{T}\right),\nonumber\\
 \sigma^\omega_{A;{\rm cl}} =& \frac{T^2}{6} + \frac{\bm{\mu}^2}{2\pi^2} +
 O\left(\frac{\mu_V \mu_A \mu_H}{T}\right), \nonumber\\
 \sigma^\omega_{H;{\rm cl}} =& 
 \frac{2\mu_V T}{\pi^2} \ln 2 + \frac{\mu_H \mu_A}{\pi^2} + 
 O\left(\frac{\mu_V \bm{\mu}^2}{T}\right). \label{eq:RKT_sigmao}
\end{align}
The circular conductivities satisfy
\begin{equation}
 \sigma^\tau_{\ell;{\rm cl}} = \frac{\mu_\ell}{6\pi^2} + O\left(\frac{\mu_V \mu_A\mu_H}{\mu_\ell T}\right).
 \label{eq:RKT_sigmat}
\end{equation}
Finally, the shear stress conductivities $\sigma^\Pi_{\ell;{\rm cl}}$ 
satisfy
\begin{align}
 \sigma^\Pi_{V;{\rm cl}} &= \frac{\mu_H}{2\pi^2 T} + O\left(\frac{\mu_H \bm{\mu}^2}{T^3}\right), \nonumber\\
 \sigma^\Pi_{A;{\rm cl}} &= \frac{1}{\pi^2} + 
 O\left(\frac{\mu_V \mu_A \mu_H}{T^3}\right), \nonumber\\
 \sigma^\Pi_{H;{\rm cl}} &= \frac{\mu_V}{2\pi^2 T} + O\left(\frac{\mu_V \bm{\mu}^2}{T^3}\right).
 \label{eq:RKT_sigmaPil}
\end{align}
\end{subequations}

Within the classical relativistic kinetic theory, the 
relations \eqref{eq:RKT_pf} giving the stress-energy tensor and 
the charge currents remain unchanged regardless of the kinematic 
state of the fluid, such that $T_{\rm cl}^{\mu\nu}$ 
and $J^\mu_{{\rm cl};\ell}$ always retain their perfect fluid form.
In Sec.~\ref{sec:QFT}, we will obtain in the frame of quantum 
field theory the corrections to the perfect fluid forms that
lead to anomalous transport. 
The classical equilibrium quantities $P_{\rm cl}$ and 
$Q_{\ell;{\rm cl}}$, as well as the coefficients 
$\sigma^\omega_{\ell;{\rm cl}}$, $\sigma^\tau_{\ell;{\rm cl}}$ 
and $\sigma^\Pi_{\ell;{\rm cl}}$, are related to thermodynamic 
integrals involving 
the Fermi-Dirac equilibrium distribution for fermions with 
the $V$, $A$ and $H$ charges and will thus serve as the basis to 
express all terms appearing as quantum corrections. 

\subsection{Kinematics of rigid motion} \label{sec:RKT:rot}

\begin{table}[t]
\begin{center}
\begin{tabular}{|c||c|c|c|c|c|c|}
\hline
 	 & $\varepsilon$ & ${\bs u}$ & ${\bs a}$ & 
 	${\bs \omega}$ & ${\bs \tau}$ \\
\hline
$C$	&   $+$	&	$+$	&	$+$ &	$+$	&	$+$\\
\hline
$P$	&   $-$	&	$-$	&	$-$ &	$+$	&	$-$\\
\hline
$T$	&   $-$	&	$-$	&	$+$ &	$-$	&	$-$\\
\hline
\end{tabular}
\end{center}
\vskip 3mm
\caption{Behaviour of the Levi-Civita symbol $\varepsilon^{\mu\nu\alpha\beta}$
and of the spatial parts of the elements of the kinematic tetrad
(velocity $\bm{u}$, acceleration $\bm{a}$, vorticity $\bm{\omega}$ and
fourth vector $\bm{\tau}$) under the $C$-, $P$-, and $T$-inversions. 
The signs $+/-$ indicate the even/odd nature of these quantities under the corresponding discrete transformations.}
\label{tbl:inversions_kinematic}
\end{table}

To reveal the role of the helical potential, we consider a gas of Dirac fermions at a finite temperature, uniformly rotating about the axis~$z$.
It is convenient to introduce a particular basis of kinematic vectors 
for the rigid motion with the four-velocity
\begin{equation}
 u^\mu \partial_\mu = \Gamma(\partial_t + \Omega \partial_\varphi), \qquad 
 \Gamma = \frac{1}{\sqrt{1 - \rho^2 \Omega^2}},\label{eq:u}
\end{equation}
where $\Omega$ is the angular velocity and $\Gamma$ is the Lorentz factor. We use cylindrical coordinates $(t,\rho, \varphi,z)$.
The acceleration $a^\mu = u^\nu \nabla_\nu u^\mu$ and vorticity 
$\omega^\mu = 
\frac{1}{2} \varepsilon^{\mu\nu\lambda\sigma} u_\nu \nabla_\lambda u_\sigma$
four-vectors are:
\begin{equation}
 a^\mu \partial_\mu =-\rho \Omega^2 \Gamma^2 \partial_\rho, \qquad 
 \omega^\mu \partial_\mu = \Omega \Gamma^2 \partial_z.
 \label{eq:omega}
\end{equation}
The fourth vector, which is orthogonal to $u$, $a$ and $\omega$, is $\tau^\mu = -\varepsilon^{\mu\nu\lambda\sigma} \omega_\nu a_\lambda u_\sigma$, or 
\begin{equation}
 \tau^\mu \partial_\mu  
 = -\Omega^3 \Gamma^5 (\rho^2 \Omega \partial_t + \partial_\varphi).
 \label{eq:tau}
\end{equation}
The properties of the elements of the kinematic tetrad under the 
$C-$, $P-$ and $T-$ inversions are summarised in
Table~\ref{tbl:inversions_kinematic}.
The squares of the vectors comprising the kinematic tetrad are
\begin{align}
 \bm{a}^2 &= -a^2 = \Omega^2 \Gamma^2(\Gamma^2 - 1), \nonumber\\
 \bm{\omega}^2 &= -\omega^2 = \Omega^2 \Gamma^4, \nonumber\\
 \tau^2 &= -\Omega^4 \Gamma^6(\Gamma^2 - 1),
\end{align}
while $u^2= 1$.

\section{Quantum field theoretical analysis}\label{sec:QFT}

In this section, we discuss the properties of the 
expectation values of the stress-energy 
tensor $T^{\mu\nu}$ (SET) and of the vector, axial and helical 
charge currents $J^\mu_{V/A/H}$, 
computed in a global equilibrium at finite temperature $T$, 
finite chemical potentials $\mu_{V/A/H}$ and finite 
rotation $\Omega$. The method employed for the computation 
relies on thermal mode sums and is summarised in 
Subsec.~\ref{sec:QFT:modes}. The results 
leading to quantum corrections to the classical results 
obtained in Sec.~\ref{sec:RKT}, as well as the new terms 
deviating from the perfect fluid form in Eq.~\eqref{eq:RKT_pf}
that give rise to vortical effects are discussed in 
Subsec.~\ref{sec:QFT:vortical}.
The computational details are given in Appendix~\ref{app:comp}.

\subsection{Thermal mode sums} \label{sec:QFT:modes}

The basis for the construction of quantum thermal states under 
rotation is the thermal average $\braket{\widehat{A}}$ of an operator 
$\widehat{A}$, defined as:
\begin{equation}
 \braket{\widehat{A}} = Z^{-1} {\rm Tr} (\hat{\varrho} \widehat{A}),
 \label{eq:tev_def}
\end{equation}
where $Z = {\rm Tr} (\hat{\varrho})$ is the partition function and 
the trace runs over Fock space~\cite{itzykson80,kapusta89,lain16,mallik16}.
We employ the density operator $\hat{\varrho}$ given by
\begin{equation}
 \hat{\varrho} = \exp\left[-\beta_0 \biggl(\widehat{H} -
 \Omega \widehat{M}^z - 
 \sum_{\ell}
 \mu_{\ell;0}\, \widehat{Q}_\ell
 \biggr)
 \right],
 \label{eq:rho}
\end{equation}
where $\widehat{H}$ is the Hamiltonian~\eq{eq:H}, 
$\widehat{M}^z = -i\partial_\varphi + S^z$ is the $z$ 
component of the total angular momentum operator, 
while $\beta_0 \equiv 1/T_0$ and $\mu_{\ell,0}$ are, respectively, 
the values of the inverse temperature and the chemical potentials 
at the rotation axis $\rho=0$ \cite{vilenkin80}. 

We now seek to evaluate the trace in Eq.~\eqref{eq:tev_def} using 
the mode sum decomposition of the field operator $\hat{\psi}$ 
given in Eq.~\eqref{eq:psi}. 
Using the fact that $[\widehat{H},\widehat{M}^z] = [\widehat{H},\widehat{P}^z] =0$, the
eigenmodes of $H$, $\gamma^5$ and $h$ introduced in Eq.~\eqref{eq:UV:chi:h} can be taken to be, simultaneously, 
the eigenmodes of $\widehat{M}^z$ and $\widehat{P}^z$:
\begin{align}
 \widehat{M}^z U_j =& m_j U_j, & 
 \widehat{M}^z V_j =& -m_j V_j,\nonumber\\
 \widehat{P}^z U_j =& k_j U_j, & 
 \widehat{P}^z V_j =& -k_j V_j,
 \label{eq:UV:M:Pz}
\end{align}
where $m_j = \pm \frac{1}{2}, \pm \frac{3}{2}, \dots$, takes 
only half-odd-integer values. These mode solutions have 
been previously derived in Ref.~\cite{ambrus14plb} and are reproduced, 
for convenience, in Eq.~\eqref{eq:modes} of Appendix~\ref{app:comp}.

The decompositions of the charge operators in Eq.~\eqref{eq:Q}, together with equivalent ones for $\cH$ and $\widehat{M}^z$ given below,
\begin{align}
 :\cH: =& \sum_j E_j(\hat{b}_j^\dagger \hat{b}_j + \hat{d}_j^\dagger \hat{d}_j), \nonumber\\
 :\widehat{M}^z: =& \sum_j m_j(\hat{b}_j^\dagger \hat{b}_j + \hat{d}_j^\dagger \hat{d}_j),
 \label{eq:HMz}
\end{align}
allow the following relations to be derived:
\begin{align}
 \hat{\varrho} \hat{b}_j^\dagger \hat{\varrho}^{-1} =& 
 e^{-\beta_0 \widetilde{E}_j - \beta(\mu_V + 2\lambda_j \mu_A + 2\lambda_j \mu_H)}
 \hat{b}^\dagger_j,\nonumber\\
 \hat{\varrho} \hat{d}^\dagger_j \hat{\varrho}^{-1} =& 
 e^{-\beta_0 \widetilde{E}_j + \beta(\mu_V - 2\lambda_j \mu_A + 2\lambda_j \mu_H)}
 \hat{d}^\dagger_j,
 \label{eq:rbr}
\end{align}
where $\widetilde{E}_j = E_j - \Omega m_j$ is the co-rotating energy.
The local inverse temperature and 
chemical potential $\beta = \Gamma^{-1} \beta_0$ and $\mu_\ell = \Gamma \mu_{\ell;0}$ are 
related to their values on the rotation axis $\beta_0$, $\mu_{\ell;0}$ via
the Lorentz factor $\Gamma$ introduced in Eq.~\eqref{eq:u}, as shown 
in Eq.~\eqref{eq:Tolman}.
Starting from Eq.~\eqref{eq:rbr}, the following t.e.v.s can be 
derived \cite{vilenkin80,mallik16}:
\begin{align}
 \braket{\hat{b}_j^\dagger \hat{b}_{j'}} =& n_{+,\lambda_j}(\tilde{p}_j)
 \delta(j,j'), \nonumber\\
 \braket{\hat{d}_j^\dagger \hat{d}_{j'}} =& n_{-,\lambda_j}(\tilde{p}_j)
 \delta(j,j'),
 \label{eq:bblocks}
\end{align}
where the notation $n_{\sigma,\lambda}(\widetilde{p})$ mimics the one considered in 
Eq.~\eqref{eq:feq}:
\begin{equation}
 n_{\sigma,\lambda}(\tilde{p}) = \frac{1}{e^{\beta_0 \tilde{p}- \beta \bm{q}_{\sigma,\lambda} \cdot \bm{\mu}} + 1}.
 \label{eq:n_def}
\end{equation}
The charges $\bm{q}_{\sigma,\lambda} \equiv (q^V_{\sigma,\lambda}, q^A_{\sigma,\lambda}, q^H_{\sigma,\lambda})$ were introduced in Eq.~\eqref{eq:q_def}.

Let us apply this formalism for the computation of the thermal expectation values (t.e.v.s) of the charge current and stress-energy 
tensor operators,
\begin{align}
 \widehat{J}^\mu_V &= \frac{1}{2} [\widehat{\overline{\psi}}, \gamma^\mu \widehat{\psi}], \qquad 
 \widehat{J}^\mu_A = \frac{1}{2} [\widehat{\overline{\psi}}, \gamma^\mu \gamma^5 \widehat{\psi}],  \nonumber\\
 \widehat{J}^\mu_H &= \frac{1}{2} [\widehat{\overline{\psi}}, \gamma^\mu h \widehat{\psi}] +
 \frac{1}{2} [\widehat{\overline{h \psi}}, \gamma^\mu \widehat{\psi}], \nonumber\\
 \widehat{T}_{\mu\nu} &= 
 \frac{i}{4} \{[\widehat{\overline{\psi}}, \gamma_{(\mu} \nabla_{\nu)} \widehat{\psi}] - 
 [\nabla_{(\mu} \widehat{\overline{\psi}} \gamma_{\nu)}, \widehat{\psi}]\}.\label{eq:J_SET_def}
\end{align}
Using the convention
$A = \braket{:\widehat{A}:}$ and the expansion \eqref{eq:psi} of the 
field operator with respect to the particle and antiparticle modes 
$U_j$ and $V_j$, Eq.~\eqref{eq:J_SET_def} can be used to 
obtain
\begin{subequations}\label{eq:J_SET_modes_aux}
\beqn J^\mu_\ell & = & \sum_j \biggl[n_{+,\lambda_j}(\tilde{p}_j) 
 \mathcal{J}^\mu_\ell(U_j, U_j) \nonumber\\
 & & - n_{-,\lambda_j}(\tilde{p}_j) 
 \mathcal{J}^\mu_\ell(V_j, V_j)\biggr],\\
 T^{\mu\nu} & = & \sum_j \biggl[n_{+,\lambda_j}(\tilde{p}_j) 
 \mathcal{T}^{\mu\nu}(U_j, U_j)  \nonumber\\
 & & - 
 n_{-,\lambda_j}(\tilde{p}_j) 
 \mathcal{T}^{\mu\nu}(V_j, V_j)\biggr].
\eeqn
\end{subequations}
The sesquilinear forms $\mathcal{J}^\mu_\ell(\psi, \chi)$ and 
$\mathcal{T}^{\mu\nu}(\psi, \chi)$ are 
\begin{align}
 \mathcal{J}^\mu_V(\psi, \chi) &= \overline{\psi} \gamma^\mu \chi, \qquad
 \mathcal{J}^\mu_A(\psi, \chi) = \overline{\psi} \gamma^\mu \gamma^5 \chi, \nonumber\\
 \mathcal{J}^\mu_H(\psi, \chi) &= \overline{\psi} \gamma^\mu h \chi + \overline{h\psi} \gamma^\mu \chi,\nonumber\\
 \mathcal{T}_{\mu\nu}(\psi,\chi) &= \frac{i}{2} 
 [\overline{\psi} \gamma_{(\mu} \nabla_{\nu)} \chi- \nabla_{(\mu} \overline{\psi} \gamma_{\nu)} \chi]
 \label{eq:sesquilinear}
\end{align}
and their explicit expressions are given in Eqs.~\eqref{eq:app_JV_sesqui} and 
\eqref{eq:app_SET_sesqui}.

The sum over $j$ appearing in Eq.~\eqref{eq:J_SET_modes_aux} 
is in principle sensitive to the choice of 
vacuum state. For the construction of rigidly-rotating thermal states,
Iyer \cite{iyer82} argued that the modes with positive co-rotating 
energy $\widetilde{E}_j = E_j - \Omega m_j > 0$ should be interpreted as
particle modes. With this convention, the modes with 
$E_j \widetilde{E}_j < 0$ will have opposite interpretation 
as compared to the stationary vacuum case. For example, a mode 
with $E_j < 0$ and $\widetilde{E}_j > 0$ will represent a 
particle with respect to the rotating vacuum and an anti-particle 
with respect to the stationary vacuum. 
The difference in results obtained with respect to the 
rotating and stationary vacua is independent of the temperature $T$ 
and chemical potentials $\mu_\ell$ of the medium and can thus be 
easily identifiable at the end of the calculation. 
We therefore present in this section the discussion in the 
simpler case of the computation 
with respect to the stationary vacuum, when
the summation over $j$ can be written as
\begin{equation}
 \sum_j \rightarrow \sum_{\lambda = \pm 1/2} 
 \sum_{m = -\infty}^\infty \int_0^\infty dp
 \, p 
 \int_{-p}^p dk.
 \label{eq:sumj_M}
\end{equation}

The results for the non-vanishing components of the charge current
can be summarised as:
\begin{multline}
 \begin{pmatrix}
  J^t_\ell \\
  \rho J^\varphi_\ell \\
  J^z_\ell
 \end{pmatrix} = 
 \sum_{m,\sigma,\lambda} 
 \frac{q^\ell_{\sigma,\lambda}}{4\pi^2}
 \int_0^\infty dp\, p\, n_{\sigma,\lambda}(\widetilde{p}) \\\times
 \int_{0}^p dk
 \begin{pmatrix}
  J_m^+ \\
  (q/p) J_m^\times \\
  2\lambda J_m^-
 \end{pmatrix},
 \label{eq:Jl_comps}
\end{multline}
where the summation runs over $m = \pm \frac{1}{2}, \pm \frac{3}{2}, \dots$, $\sigma = \pm 1$ and $\lambda = \pm \frac{1}{2}$. The functions
$J_m^* \equiv J_m^*(q\rho)$ ($* \in \{+,-,\times\}$) are 
quadratic with respect to the Bessel functions,
\begin{align}
 J_m^\pm \equiv J_m^\pm(q\rho) &= J_{m-\frac{1}{2}}^2(q\rho) - 
 J_{m+\frac{1}{2}}^2(q\rho), \nonumber\\ 
 J_m^\times \equiv J_m^\times(q\rho) &= 2J_{m-\frac{1}{2}}(q\rho) 
 J_{m+\frac{1}{2}}(q\rho).
\end{align}
The components of the stress-energy tensor are
\begin{multline}
 \begin{pmatrix}
  T^{tt} \\
  \rho T^{t\varphi} \\
  T^{tz} \\
  \rho^2 T^{\varphi\varphi} \\
  \rho T^{\varphi z} \\
  T^{zz}
 \end{pmatrix} = \sum_{m,\sigma,\lambda} 
 \frac{1}{4\pi^2}
 \int_0^\infty dp\, p\, n_{\sigma,\lambda}(\widetilde{p})
 \\\times \int_{0}^p dk 
 \begin{pmatrix}
  p J_m^+ \\ 
  \frac{m}{2\rho} J_m^+ - \frac{1}{4\rho} J_m^- + \frac{q}{2} J_m^\times, \\
  \frac{\lambda}{p}(p^2 + k^2) J_m^- \\
  \frac{q m}{\rho p} J_m^\times \\
  \frac{\lambda}{\rho}(m J_m^- - \frac{1}{2} J_m^+) \\
  \frac{k^2}{p} J_m^+
 \end{pmatrix},
 \label{eq:SET_comps}
\end{multline}
while $T^{\rho\rho} = T^{zz}$.

\subsection{Vortical effects in V/A/H fluids}\label{sec:QFT:vortical}


\begin{table*}[t]
\centering
\begin{tabular}{|c||c|c|c||c|c|c||c|c|c||c|c|c|c||}
        \hline
         & 
         $\sigma^\omega_V$ & $\sigma^\omega_A$ & $\sigma^\omega_H$ & 
         $\sigma^\tau_V$ & $\sigma^\tau_A$ & $\sigma^\tau_H$ &
         $\sigma^\Pi_V$ & $\sigma^\Pi_A$ & $\sigma^\Pi_H$ &
         $\sigma^\omega_\varepsilon$ & $\sigma^\tau_\varepsilon$ & $\Pi_1$ & $\Pi_2$ \\\hline
         $C$ & $-$ & $+$ & $-$ &
         $-$ & $+$ & $-$ & 
         $-$ & $+$ & $-$ &
         $+$ & $+$ & $+$ & $+$ \\\hline 
         $P$ & $-$ & $+$ & $+$ &
         $+$ & $-$ & $-$ & 
         $-$ & $+$ & $+$ &
         $-$ & $+$ & $+$ & $-$ \\\hline 
         $T$ & $+$ & $+$ & $+$ & 
         $+$ & $+$ & $+$ & 
         $+$ & $+$ & $+$ &
         $+$ & $+$ & $+$ & $+$ \\\hline
\end{tabular}
\vskip 3mm
\caption{CPT properties of the vortical and circular conductivities $\sigma^\omega_\ell$ and $\sigma^\tau_\ell$ appearing in 
Eq.~\eqref{eq:currents}. The shear charge conductivities 
$\sigma^\Pi_\ell$ are also shown for completeness. 
The last four columns show the vortical and circular heat
conductivities $\sigma^\omega_\varepsilon$ and $\sigma^\tau_\varepsilon$,
together with the shear coefficients $\Pi_1$ and $\Pi_2$ appearing in 
Eq.~\eqref{eq:PiW}. The $T$-even nature of all transport coefficients in this table highlights the dissipationless nature of the transport phenomena, including, in particular, the new transport related to the helical degrees of freedom.}
\label{tbl:CPT_coeffs}
\end{table*}

The rigidly rotating gas of Dirac fermions generates the vector, axial, and helical 4-currents $(\ell=V,A,H)$ according to Eq.~\eqref{eq:Jl_comps}, which can be rearranged as follows:
\begin{equation}
J^\mu_\ell \equiv \braket{:\widehat{J}^\mu_\ell:} = 
 Q_\ell u^\mu + \sigma_\ell^\omega \omega^\mu + \sigma_\ell^\tau \tau^\mu,
\label{eq:currents}
\end{equation}
along the four-vectors $u^\mu$, $\omega^\mu$ and $\tau^\mu$. The radial components along $a^\mu$ are absent for all currents~\eqref{eq:currents}. Henceforth, we work in the $\beta$ (thermometer) frame, by fixing the four-velocity $u^\mu$ to be equal to the one given in Eq.~\eqref{eq:u} \cite{van12,van13,becattini15epjc}. Combining Tables~\ref{tbl:inversions} and \ref{tbl:inversions_kinematic},
the CPT parities of $\sigma^\omega_\ell$ and $\sigma^\tau_\ell$ 
can be obtained, as summarised in Table~\ref{tbl:CPT_coeffs}. 
Looking now at the thermodynamic relations proposed in
Eq.~\eqref{eq:RKT_sigmas_P}, it can be seen that the CPT properties 
of $\sigma^\omega_\ell$ and $\sigma^\tau_\ell$ should be the same as 
those of $Q_\ell Q_A$ and $Q_\ell$, respectively, which is easily 
confirmed by comparing Tables~\ref{tbl:inversions} and 
\ref{tbl:CPT_coeffs}.
Table \ref{tbl:CPT_coeffs} highlights, in particular, that helicity and chirality are different quantities that should not be confused with each other~\cite{peskin95}. 

The thermal expectation value of the stress energy tensor $T_{\mu\nu}$ admits the 
general decomposition with respect to $u^\mu$ as follows:
\begin{equation}
 T^{\mu\nu} = (E + P) u^\mu u^\nu - P g^{\mu\nu} + \Pi^{\mu\nu} + W^\mu u^\nu + u^\mu W^\nu,
 \label{eq:SET}
\end{equation}
where $P$ is the isotropic pressure and $E = 3P$ is the energy density. 
Both the anisotropic stress $\Pi^{\mu\nu}$ and the heat flux $W^\mu$ are orthogonal to $u^\mu$ and in addition, $\Pi^{\mu\nu}$
is required to be traceless.
The most general decomposition of 
$\Pi^{\mu\nu}$ and $W^\mu$ compatible with the 
tensor 
structure of $T^{\mu\nu}$ exhibited in Eq.~\eqref{eq:SET_comps} is
\begin{align}
 \Pi^{\mu\nu} &= 
 \Pi_1\left(\tau^\mu \tau^\nu - \frac{\bm{\omega}^2}{2} a^\mu a^\nu -
 \frac{\bm{a}^2}{2} \omega^\mu \omega^\nu \right) \nonumber\\
 & + \Pi_2(\tau^\mu \omega^\nu + \tau^\nu \omega^\mu), \nonumber\\
 W^\mu &= \sigma_\varepsilon^\tau \tau^\mu + \sigma_\varepsilon^\omega \omega^\mu.\label{eq:PiW}
\end{align}
The properties of the transport coefficients $\sigma^\omega_\varepsilon$, 
$\sigma^\tau_\varepsilon$, $\Pi_1$ and $\Pi_2$ under the $C-$, $P-$ and 
$T-$ inversions, summarised in Table~\ref{tbl:CPT_coeffs}, 
can be obtained based on those of the kinematic vectors
$u^\mu$, $a^\mu$, $\omega^\mu$ and $\tau^\mu$, given in 
Table~\ref{tbl:inversions_kinematic}.

We now report the results for the terms appearing in the decompositions 
of the charge currents and of the stress-energy current:
\begin{align}
 P &= P_{\rm cl} + \frac{3\bm{\omega}^2 + \bm{a}^2}{12} \sigma^\omega_{A;{\rm cl}} \nonumber \\
& 
 + \frac{45\bm{\omega}^4 + 46\bm{\omega}^2 \bm{a}^2- 51\bm{a}^4}{8640} \sigma^\Pi_{A;{\rm cl}} + \Delta P^{(h)}, \nonumber\\
 Q_\ell &= Q_{\ell; {\rm cl}} + \frac{3(\bm{\omega}^2 + \bm{a}^2)}{2} 
 \sigma^\tau_{\ell;{\rm cl}} + Q_\ell^{(h)}, \nonumber\\
 \sigma^\omega_\ell &= \sigma^\omega_{\ell,{\rm cl}} + 
 \frac{\bm{\omega}^2 + 3 \bm{a}^2}{24} \sigma^\Pi_{\ell;{\rm cl}} + \sigma^{\omega, (h)}_\ell,\nonumber\\
 \sigma^\tau_\ell &= \sigma^\tau_{\ell;{\rm cl}} + \sigma^{\tau,(h)}_\ell,\nonumber\\
 \sigma^\omega_\varepsilon &= Q_{A;{\rm cl}} + \frac{\bm{\omega}^2 + \bm{a}^2}{2} \sigma^{\tau}_{A;{\rm cl}} + \sigma^{\omega,(h)}_\varepsilon, \nonumber\\
 \sigma^\tau_\varepsilon &= -\frac{1}{3} \sigma^\omega_{A;{\rm cl}} -
 \frac{39 \bm{\omega}^2 + 31\bm{a}^2}{360} \sigma^\Pi_{A;{\rm cl}} + 
 \sigma^{\tau,(h)}_{\varepsilon},\nonumber\\
 \Pi_1 &= -\frac{2}{27} \sigma^\Pi_{A;{\rm cl}} + \Pi_1^{(h)}, \quad 
 \Pi_2 = -2 \sigma^\tau_{A;{\rm cl}} + \Pi_2^{(h)},
 \label{eq:QFT}
\end{align}
where the quantities bearing the ``${\rm cl}$'' subscript are defined 
in Eq.~\eqref{eq:RKT_Li} [their high temperature expansion can be 
found in Eq.~\eqref{eq:RKT_highT}].
The terms with the ``$(h)$'' superscript represent higher-order, subleading corrections,
summarised in Appendix~\ref{app:comp},
along with further computational details.

The above expressions are fully consistent (and indeed, constrained) 
by the CPT properties revealed in Tables~\ref{tbl:inversions} and 
\ref{tbl:CPT_coeffs}. These restrictions are even more important 
in determining the coefficients appearing on the SET sector. For example, the pressure has even parity ($+++$)
with respect to all discrete transformations and can therefore 
receive corrections proportional only to $\sigma^\omega_{A;{\rm cl}}$ and
$\sigma^\Pi_{A;{\rm cl}}$. 
Since $\sigma^\omega_\varepsilon$ has units of $(\text{energy})^3$ and 
CPT parity equal to $(+-+)$, its leading order contribution must be 
proportional (in fact, equal) to $Q_{A;{\rm cl}}$, 
while its first correction of order $(\text{energy})^1$
is necessarily proportional to $\sigma^\tau_{A;{\rm cl}}$. Conversely, since 
$\sigma^\tau_\varepsilon$ has dimensions $(\text{energy})^2$ and parity 
$(+++)$, its leading order contribution is proportional to 
$\sigma^\omega_{A; {\rm cl}}$, while its dimensionless correction 
is proportional to $\sigma^\Pi_{A;{\rm cl}}$, just like the leading
order contribution to $\Pi_1$. Finally, the dimensionality and $(+-+)$
parity of $\Pi_2$ indicate that its leading order term must be 
related to $\sigma^\tau_{A;{\rm cl}}$.

The vortical transport effects~\eqref{eq:currents} and \eqref{eq:PiW} are
consistent with the $C$-, $P$-, and $T$-symmetries of the vector, axial 
and helical currents and charges, as shown in Table~\ref{tbl:inversions}. 
These are dissipationless effects because the laws
\eqref{eq:currents}--\eqref{eq:PiW} are even under the $T$-inversion. 

On the rotation axis ($\rho = 0$), the circular currents vanish and the
currents~\eqref{eq:currents} and heat flux \eqref{eq:PiW} point exactly 
along the vorticity $\bm{\Omega}$. The rotating dense (charged) Dirac 
matter generates on the rotation axis the helical current 
${\bs J}_H = \sigma^\omega_H \bm{\omega}$ 
that is linearly proportional to the vector chemical potential $\mu_V$ 
and temperature $T$~\eqref{eq:RKT_sigmao}. On the other hand, the neutral 
Dirac matter with nonzero helicity ($\mu_H \neq 0$) generates the vector 
(charge) current~${\bs J}_V = \sigma^\omega_V \bm{\omega}$. Remarkably, 
the mentioned helical terms, linearly proportional to a chemical potential 
and temperature, are allowed for the helical effects and, at the same time, 
are forbidden for the chiral phenomena by virtue of the $C$-, $P$-, and
$T$-symmetries, as pointed out above.

\section{Applications and discussion}\label{sec:app}

Let us now consider how the helical degrees of freedom reveal themselves in an interacting field theory. 
In the previous section, we pointed out that finite vorticity gives rise 
to transport related to the helical current and the corresponding helical 
chemical potential. This allows us to speculate that in the underlying 
(fundamental) interacting theory (such as QED), triangle diagrams 
involving the helical vertex may be anomalous similarly to the well-known anomalous 
triangular diagrams which incorporate the axial current. We address this issue 
in Subsec.~\ref{sec:app:anomalies}. 
In Subsec.~\ref{sec:app:HVW}, we point out that 
wave-like excitations involving both vector and helical currents appear in neutral Dirac materials in the presence of vorticity.
Finally, we discuss in Subsec.~\ref{sec:app:condmat} 
potential implications of dynamical helical degrees of freedom for condensed matter systems 
by pointing out the possibility of whirlpool structures emerging 
in Dirac fluids.

\subsection{Helical anomalies in QED}\label{sec:app:anomalies}

Fermions with electric charge $e$ couple to electromagnetism through a source term $e J^\mu_V A_\mu$ added to the Lagrangian~\eq{eq:L}: $\cL \to \cL + e J^\mu_V A_\mu$. The $U(1)_V$ symmetry~\eq{eq:U1:V} is unbroken at the quantum level, thus reflecting the fundamental property of the electric charge conservation in quantum theory.
The axial symmetry~\eq{eq:U1:A}, however, is broken at the quantum level, leading to the nonconservation of 
$J^\mu_A$ in the presence of the electric $\bs E$ and magnetic $\bs B$ background fields~\cite{FujikawaBook}:
\beqn
\partial_\mu J^\mu_A = 
-\frac{e^2}{8\pi^2} F_{\mu\nu} {\widetilde F}^{\mu\nu}
\equiv \frac{e^2}{2\pi^2}{\bs E} \cdot  {\bs B},
\label{eq:dJ:A}
\eeqn
where ${\widetilde F}^{\mu\nu} = \frac{1}{2} \varepsilon^{\mu\nu\alpha\beta}F_{\alpha\beta}$ and $F_{\mu\nu} = \partial_\mu A_\nu - \partial_\nu A_\mu$.

While elastic processes due to the electromagnetic interaction conserve both chirality and helicity \cite{itzykson80,peskin95,Kapusta20}, the inelastic processes of the form pair annihilation $\rightarrow$ pair creation may lead to helicity non-conservation. For example, in the process $e^+_R e^-_L \rightarrow e^+_L e^-_R$, the axial charge of the initial and final states is $0$, thus being conserved. The helicity charge of the initial state is $-2$ (here $e^-_L$ is a particle with negative helicity), while in the final state, it is $+2$. We will discuss the efficiency of such processes in destroying helicity imbalance in the next subsection.

It is known that the vector~$\sigma^\omega_V$ and axial~$\sigma^\omega_A$ vortical conductivities~\eqref{eq:RKT_sigmao} at $\mu_H = 0$ are determined by the axial quantum anomalies~\cite{Landsteiner:2012kd}. For example, the $\mu_V\mu_A$ term in $\sigma^\omega_V$ is generated by the axial-vector-vector ($AVV$) vertex of the axial anomaly~\eqref{eq:dJ:A}, which is also responsible for the $\mu_V^2$ term in $\sigma^\omega_A$. Both these terms share similar prefactors with the axial anomaly~\eqref{eq:dJ:A}. The axial conductivity $\sigma^\omega_A$ contains also the $\mu_A^2$ term due to the axial-axial-axial ($AAA$) triangular anomaly, as well as the $T^2$ term which originates from the axial-graviton-graviton ($ATT$) vertex of the mixed axial-gravitational anomaly~\cite{Landsteiner:2011cp}. 

The presence of the helical component in the vortical conductivities~\eqref{eq:RKT_sigmao} strongly suggests the existence of new types of triangle anomalies which involve the helicity vertex~\eq{eq:h}. The new helical anomalies must reveal themselves in the background of a ``helical vector field'' $A^H_\mu$ which couples with the Dirac fermions via the source term $A^H_\mu J^\mu_H$ added to the Lagrangian~\eq{eq:L}. For instance, the leading $\mu_H T$ ($\mu_V T$) terms  in $\sigma^\omega_V$ ($\sigma^\omega_H$), as well as the $\mu_V \mu_H T$ term in $\sigma_\varepsilon^\omega$, could have their origin in the new triangle $HTV$ anomaly involving vector ``$V$'' ($\gamma^\mu$), helical ``$H$'' ($\gamma^\mu h$), and graviton ``$T$'' (${\widehat T}^{\mu\nu}$) vertices. 
The quadratic $\mu_H$ dependence~\eq{eq:RKT_sigmao} of 
$\sigma^\omega_A$ implies the existence of a particular form of the mixed axial-helical anomaly responsible for the nonconservation of the axial current 
via a $AHH$ vertex that shares similarity with the standard $AVV$ vertex of the axial anomaly~\eq{eq:dJ:A}, to be discussed in Ref.~\cite{ref:in:preparation}.

\subsection{Helicity relaxation time}\label{sec:app:tauH}

In QED, helicity conservation is violated in the pair annihilation processes of the form $e^+_R e^-_L \rightarrow e^+_L e^-_R$. The differential cross section corresponding to such helicity-violating pair-annihilation (HVPA) processes is \cite{peskin95}
\begin{equation}
 \frac{d\sigma}{d\Omega}(e^+_R e^-_L \rightarrow e^+_L e^-_R) = \frac{\alpha^2}{4E_{\rm cm}^2} (1 - \cos\theta_{cm})^2, 
 \label{eq:crosss_QED}
\end{equation}
where $\alpha$ is the fine-structure constant. The center-of-momentum energy $E_{cm}$ and scattering angle $\theta_{cm}$ are related to the Mandelstam variables $s$ and $t$ via 
\begin{align}
 s &= (p + k)^2 = E_{cm}^2, \nonumber\\
 t &= (p - p')^2 = -\frac{1}{2} E_{cm}^2(1 - \cos\theta_{cm}),
\end{align}
where $(p,k)$ and $(p',k')$ are the incoming and outgoing momenta, respectively.

In the QGP, the relevant fermionic degrees of freedom are the (massless) quarks and the HVPA processes take place via gluon exchange. Eq.~\eqref{eq:crosss_QED} is modified by replacing $\alpha$ by $\alpha_{QCD}$ and by including the $SU(3)$ generators from the gluon vertices~\cite{peskin95}:
\begin{multline}
 \frac{d\sigma}{d\Omega}(q^{i}_R \bar{q}^{j}_L \rightarrow q^{i'}_L \bar{q}^{j'}_R) = \frac{\alpha_{QCD}^2}{4E_{\rm cm}^2} (1 - \cos\theta_{cm})^2 \\ 
 \times \sum_{a,b} t^a_{ji} t^a_{j'i'} t^b_{ij} t^b_{i'j'}, 
 \label{eq:crosss_QCD}
\end{multline}
where $t^a_{ij}$ ($1 \le a \le 8$) are the SU(3) generators in the fundamental representation, normalized such that ${\rm tr}(t^a t^b) = \frac{1}{2} \delta^{ab}$, while $(i,j)$ and $(i',j')$ are the colour indices of the initial and final quarks. Note that in Eq.~\eqref{eq:crosss_QCD}, we are neglecting higher-order diagrams such as gluon emission or gluon exchange between the outgoing quarks. 

Let us estimate the helicity relaxation time corresponding to an infinitesimal but small helical density. The Boltzmann equation for the distribution function of fermions ($\sigma = 1$) / anti-fermions ($\sigma = -1$) with on-shell four-momentum $p^\mu = (E_\bp, \bp)$ and helicity $\lambda$ reads
\begin{equation}
p^\mu \partial_\mu f^\sigma_{\mathbf{p},\lambda} = C[f],\label{eq:boltz}
\end{equation}
where $C[f]$ collects all processes in which such fermions are involved. The helical charge can be computed via Eq.~\eqref{eq:RKT_P_Q}, reproduced below for convenience:
\begin{equation}
 Q_H = g \sum_{\sigma,\lambda} 2\lambda \sigma \int dP (p \cdot u) f^\sigma_{\bp, \lambda},
\end{equation}
where $dP = d^3p / [(2\pi)^3 E_\bp]$, $u^\mu$ is the local plasma four-velocity,
while $\sigma$ and $\lambda$ take the values $\pm 1$ and $\pm 1/2$, respectively. The degeneracy factor $g = N_c N_f$ accounts for $N_c = 3$ QCD colours and $N_f$ quark flavours.

The time derivative of $Q_H$ in a spatially-homogeneous system at rest will thus be given by the helicity-violating pair-annihilation (HVPA) processes contribution to $C[f]$ \cite{Manuel:2015zpa,Ruggieri:2016asg},
\begin{equation}
 \frac{dQ_H}{dt} = g \sum_{\lambda, \sigma} 2 \sigma \lambda \int dP C_{\mathrm{HVPA}}[f],
 \label{eq:dQH_gen}
\end{equation}
where $C_{\mathrm{HVPA}}[f]$ is modeled via \cite{DeGroot:1980dk}
\begin{multline}
 C_{\mathrm{HVPA}}[f] = 
 \int dK dP' dK' \delta^4(p + k - p' -k') 
 s (2\pi)^6
 \\\times  
 [f_{\bp',-\lambda}^\sigma f^{-\sigma}_{\bk',\lambda} \tilde{f}^\sigma_{\bp,\lambda} \tilde{f}^{-\sigma}_{\bk, -\lambda}
 - f_{\bp,\lambda}^\sigma f^{-\sigma}_{\bk,-\lambda} \tilde{f}^\sigma_{\bp',-\lambda} \tilde{f}^{-\sigma}_{\bk', \lambda}] \\
 \times N_f \sum_{i',j,j'} \frac{d\sigma}{d\Omega}(q^{\sigma,i}_{\bp,\lambda} q^{-\sigma,j}_{\bk,-\lambda} \rightarrow q^{\sigma,i'}_{\bp',-\lambda} q^{-\sigma,j'}_{\bk',\lambda}).
 \label{eq:C_HVPA}
\end{multline}
where the $\tilde{f}^\sigma_{\bp,\lambda} = 1 - f^\sigma_{\bp,\lambda}$ factors originate from the Pauli blocking for fermions. The first (second) term in the square bracket is the gain (loss) term. 
The variables $(p^\mu, k^\mu)$ and $(p'{}^\mu, k'{}^\mu)$ denote the momenta of the outgoing (incoming) and incoming (outgoing) particles in the gain (loss) term. 
Focusing on the loss term, a quark of type $\sigma$ with momentum $\bp$ and helicity $\lambda$, having colour $i$ and flavour $f$, annihilates with a quark of type $-\sigma$, with momentum $\bk$, colour $j$ and opposite helicity $-\lambda$, but with the same flavour $f$. Under the HVPA process, the emerging quark with type $\sigma$ has flipped helicity, $-\lambda$, colour $i'$ and flavour $f'$. Its partner of type $-\sigma$ will have helicity $\lambda$, colour $j'$ and the same flavour $f'$. The arbitrariness of $f'$ is accounted for by the $N_f$ factor on the last line of Eq.~\eqref{eq:C_HVPA}, while the colours $i'$, $j$ and $j'$ can be summed starting from:
\begin{align}
 \sum_{a,b} \sum_{i',j,j'} t^a_{i'i} t^a_{j'j} t^b_{ii'} t^b_{jj'} = \sum_{a,b} (t^b t^a)_{ii} 
 {\rm tr}(t^b t^a). 
\end{align}
Taking into account that ${\rm tr}(t^b t^a) = \frac{1}{2} \delta^{ab}$ and that $\sum_a t^a t^a = \frac{4}{3} \mathbb{I}_{3\times 3}$, the above sums evaluate to $2/3$, such that 
\begin{multline}
 N_f \sum_{i',j,j'} \frac{d\sigma}{d\Omega}(q^{\sigma,i}_{\bp,\lambda} q^{-\sigma,j}_{\bk,-\lambda} \rightarrow q^{\sigma,i'}_{\bp',-\lambda} q^{-\sigma,j'}_{\bk',\lambda}) \\ 
 = g \frac{\alpha_{QCD}^2}{18E_{\rm cm}^2} (1 - \cos\theta_{cm})^2.
\end{multline}

We consider now a near-equilibrium neutral plasma with a small helical imbalance, such that 
\begin{equation}
 f^\sigma_{\bp,\lambda} \simeq f_{0\bp} + 2\lambda \sigma \beta \mu_H f_{0\bp} \tilde{f}_{0\bp},
\end{equation}
with $f_{0\bp} = [e^{\beta E_\bp} + 1]^{-1}$ being the Fermi-Dirac distribution for a neutral, unpolarized fluid. Ignoring quadratic terms in $\mu_H$, the sum over $\lambda$ and $\sigma$ can be performed in Eq.~\eqref{eq:dQH_gen}, leading to
\begin{multline}
 \frac{dQ_H}{dt} = -16 \beta \mu_H g^2 \int dP dK dP' dK' 
 s (2\pi)^6 \\\times 
 \delta^4(p + k - p' -k') 
 f_{0\bp}  f_{0\bk} \tilde{f}_{0\bp'} \tilde{f}_{0\bk'}(\tilde{f}_{0\bp} + f_{0\bp'}) \\ 
 \times \frac{\alpha_{QCD}^2}{18E_{\rm cm}^2} (1 - \cos\theta_{cm})^2.
 \label{eq:dQH_summed}
\end{multline}
Writing  
\begin{equation}
\frac{dQ_H}{dt} = -\frac{Q_H}{\tau_H},
\end{equation}
with $Q_H = g \mu_H / 3 \beta^2$,
the relaxation time $\tau_H$ can be evaluated via
\begin{multline}
 \tau_H^{-1} = \frac{8}{3} (2\pi)^6 g \alpha_{\rm QCD}^2 \beta^3 \int dP dK dP' dK' \\\times (1 - \cos\theta_{cm})^2
 \delta^4(p + k - p' -k')  
 \\\times 
 f_{0\bp}  f_{0\bk} \tilde{f}_{0\bp'} \tilde{f}_{0\bk'}(\tilde{f}_{0\bp} + f_{0\bp'}).
 \label{eq:tauH_inv}
\end{multline}
The above integral can be evaluated in the CM frame (the details can be found in Appendix~\ref{app:tauH}), giving 
\begin{align}
 \tau_H &= 0.208 \times \frac{\pi^3 \beta}{N_f \alpha_{\rm QCD}^2} \nonumber\\
 &\simeq \left(\frac{250\ {\rm MeV}}{k_B T}\right) \left(\frac{1}{\alpha_{\rm QCD}}\right)^2 \left(\frac{2}{N_f}\right) \nonumber \\
 & \times 2.54 \ {\rm fm}/c.
 \label{eq:tauH_estimate}
\end{align}
For two light fermionic flavors, $N_f = 2$, the QGP in the strongly coupled ($\alpha_{\rm QCD} \simeq 1$) regime above the crossover ($T = 250 \, \mathrm{MeV}$) gives us, theoretically:
\begin{align}
    \tau_H \simeq 2.54\,{\rm fm}/c\,.
\end{align}

On the other hand, estimations of the relaxation rate of the axial charge~$\tau_A$ from first-principle simulations~\cite{Astrakhantsev:2019zkr}
give 
\begin{align}
    \tau_A \simeq 0.25\,\mathrm{fm}/c\,, 
\end{align}
at temperatures above the QCD crossover. An estimate based on the NJL model \cite{Ruggieri:2016asg} gives the axial relaxation time in the range $\tau_A \simeq (0.1 \dots 1)\ {\rm fm}/c$.

The relaxation times $\tau_A$ and $\tau_H$ are lower than (or comparable with) the typical lifetime for the QGP, which amounts to several ${\rm fm}/c$, indicating that non-elastic processes are rather effective in altering the helicity imbalance in the context of heavy ion collisions. Therefore, our crude estimations show that the helical charge decays is comparable or slower than that of the axial charge, $\tau_H \gtrsim \tau_A$, implying that both of them should be treated at the same level.

\subsection{Helical vortical waves}\label{sec:app:HVW}

The emergence of the helicity degree of freedom allows us to uncover new hydrodynamic excitations in the helical sector, that are similar to the chiral magnetic~\cite{Kharzeev:2010gd} and chiral vortical~\cite{Jiang:2015cva} waves. To illustrate this fact, we neglect background electromagnetic fields and impose the conservation of the charge currents $J^\mu_\ell$ in Eq.~\eqref{eq:currents} for the case of a rigidly-rotating fluid, when $u^\mu$, $\omega^\mu$ and $\tau^\mu$ are those given in Eqs.~\eqref{eq:u}, \eqref{eq:omega} and \eqref{eq:tau}, respectively. Noting that $\partial_\mu u^\mu = \partial_\mu \omega^\mu = \partial_\mu \tau^\mu = 0$, the 
divergences $\partial_\mu J^\mu_\ell$ of the vector, axial and helical currents
can be evaluated on the $z$ axis (when $\rho = 0$), reducing to
\begin{align}
 \partial_t Q_V + \Omega \partial_z \sigma^\omega_V &= 0,\nonumber\\
 \partial_t Q_A + \Omega \partial_z \sigma^\omega_A &= -\frac{Q_A}{\tau_A},\nonumber\\
 \partial_t Q_H + \Omega \partial_z \sigma^\omega_H &= -\frac{Q_H}{\tau_H},
 \label{eq:hvw_cons}
\end{align}
where we took into account that neither the axial nor the helical charge remain conserved in a strongly-interacting plasma.

We take a globally neutral plasma ($\overline{Q}_\ell = 0$, $\ell=V,A,H$) at finite temperature ($T \neq 0$), and consider the simplest 
excitation that propagates along the vorticity vector on the rotational axis, where the mean fluid velocity vanishes, ${\bar {\bs v}} = 0$. The bar over a symbol means a local thermodynamic average.
We consider linear modes in hydrodynamic fluctuations. Neglecting the quantum corrections such that $(Q_\ell,\sigma^\omega_\ell) \simeq (Q_{\ell;{\rm cl}}, \sigma^{\omega}_{\ell;{\rm cl}})$, computed in Eqs.~\eqref{eq:RKT_Ql}--\eqref{eq:RKT_sigmao}, we have
\begin{gather}
 \delta Q_\ell \simeq \frac{T^2}{3} \delta \mu_\ell,  \quad \delta \sigma^\omega_A \simeq \frac{T}{3} \delta T, \nonumber\\
 \delta \sigma^\omega_{V/H} \simeq \frac{2T \ln 2}{\pi^2} \delta \mu_{H/V}. \label{eq:hvw_linear}
\end{gather}
Plugging the above into Eq.~\eqref{eq:hvw_cons}, it can be seen that the fluctuations in the axial chemical potential, $\delta \mu_A$, are related to the fluctuations in temperature, $\delta T$, which are in principle governed by the hydrodynamic equations following from $\nabla_\mu T^{\mu\nu} = 0$. For this 
reason, the chiral vortical wave (CVW) -- that involves the vector and axial sectors -- does not propagate in the neutral plasma~\cite{Kalaydzhyan:2016dyr,Abbasi:2016rds,footnote:3,Chernodub:2015gxa}. Unexpectedly, the vector and helical chemical potentials are linearly cross-coupled and therefore the helical vortical wave (HVW) does propagate. This can be seen by substituting $(\delta Q_{V/H}, \sigma^\omega_{V/H})$ given in Eq.~\eqref{eq:hvw_linear} into Eq.~\eqref{eq:hvw_cons}, leading to
\begin{align}
 \partial_t \delta Q_{V} + \frac{6\Omega \ln 2}{\pi^2 T} \partial_z \delta Q_{H} &= 0,\nonumber\\
 \partial_t \delta Q_{H} + \frac{6\Omega \ln 2}{\pi^2 T} \partial_z \delta Q_{V} &= -\frac{\delta Q_H}{\tau_H},
\end{align}
where $\tau_H$ is the helicity relaxation time discussed in Sec.~\ref{sec:app:tauH}.
Combining the above relations leads to the wave equation
\beqn
(\partial_t^2 - \tau_H^{-1} \partial_t - v_{\HVW}^2 \partial_z^2) \delta Q_H = 0\,,
\eeqn
where the HVW velocity $v_{\HVW}$ satisfies (we restore the fundamental constants $\hbar$, $k_B$, and $c$)
\beqn
v_{\HVW} = \frac{6 \ln 2}{\pi^2} \, \frac{\hbar \lvert\Omega\rvert}{k_B T} c.
\label{eq:v:HVW}
\eeqn

Let us critically assess the possibility of emergence of the HVW in the realistic environment of heavy-ion collisions. To estimate the velocity of the HVW in ultra-relativistic heavy-ion collisions, we take the temperature $T \simeq 250\,\mathrm{MeV}$ above the pseudocritical QCD value~\cite{Jacak:2012dx,Aoki:2006we}, and the vorticity $\Omega \simeq 6.6\,\mathrm{MeV}\simeq 10^{22} \, s^{-1}$ revealed in a RHIC experiment~\cite{STAR:2017ckg,Wang:2017jpl}. We find that at these parameters, the HVW propagates with the velocity $v_{\HVW} \simeq 1 \times 10^{-2} c$, such that 
\begin{equation}
 v_{\HVW}  \simeq 10^{-2} c\, \left(\frac{\lvert \hbar \Omega \rvert}{6.6\ {\rm MeV}}\right) \left(\frac{250\ {\rm MeV}}{k_B T}\right).
\end{equation}
Looking now at the full dispersion relation for Fourier modes with angular velocity $\omega$ and wavenumber $k$,
\begin{equation}
 \omega_\pm = -\frac{i}{2\tau_H} \pm k v_{\HVW} \sqrt{1 - \frac{1}{4 k^2 v_{\HVW}^2 \tau_H^2}}.
\label{eq:omega:standard}
\end{equation}
We see that the HVW propagates only for sufficiently high wavenumbers, $k v_{\HVW} > 1 / (2\tau_H)$. For wavelength $\lambda \equiv 2\pi/k$, this condition transforms to the requirement $\lambda < 4\pi l_H$ implying the wavelength of the HVW should be smaller than $4\pi$ times
the distance $l_H = v_{\HVW} \tau_H$ the wave propagates during the helicity-relaxation time $\tau_H$. In our case, $l_H \simeq 2.5 \times 10^{-2} \mathrm{fm}$, so that the propagating waves should have rather short wavelength $\lambda \lesssim 
0.3\, \mathrm{fm}$, or as a function of temperature, 
\begin{multline}
 \lambda < \left(\frac{\hbar \Omega}{6,6\ {\rm MeV}}\right) \left(\frac{250\ {\rm MeV}}{k_B T \alpha_{\rm QCD}}\right)^2 \left(\frac{2}{N_f}\right) \\\times 0.3\ {\rm fm}.
\end{multline}
At $T = 250\ {\rm MeV}$, 
the lowest wavenumber bound is $k \gtrsim 20\, \mathrm{fm}^{-1}$. However, despite the shortness of the wavelength of the propagating helical vortical wave, its energy scale, given by the real part of the frequency~\eq{eq:omega:standard}, is a rather modest quantity, $\varepsilon_{\HVW} \equiv \hbar k v_{\HVW} \gtrsim 0.2 \hbar c \cdot {\mathrm{fm}}^{-1} \approx 40 \,{\mathrm{MeV}}$, owing to the slow propagation velocity $v_{\HVW}$. With respect to the local temperature, the HVW energy must satisfy
\begin{equation}
 \varepsilon = k v_{\HVW} > \frac{1}{2\tau_H} \simeq 0.16 k_B T\,,
\end{equation}
so that the propagating HVW may be thermally excited in the QGP. The mode is not gapless in the sense that the lowest wavenumber is bounded from below.
We note that the lower bound for the energy of the propagating HVW was computed based on our perturbative estimation of the helicity relaxation time. A more accurate estimate requires lattice QCD techniques \cite{Astrakhantsev:2019zkr} and lies beyond the scope of our work.

It is instructive to compare the parameters of the HVW~\eqref{eq:v:HVW} with its chiral analog~\cite{Jiang:2015cva,Kalaydzhyan:2016dyr}:
\begin{align}
{v}_{\CVW} = \frac{3 \mu_V \Omega}{\pi^2 T^2}\,.
\end{align}
To this end, we set for the chemical potential $\mu_V = \mu_q \simeq 30\,\mathrm{MeV}$~\cite{Huang:2011ez}, and obtain the velocity of the CVW, $v_{\CVW} \simeq 1 \times 10^{-3} c$, which falls into a range of the original estimation of Ref.~\cite{Jiang:2015cva}. The HVW propagates much faster than the CVW: 
\beqn
\frac{v_{\HVW}}{v_{\CVW}} = 2 \ln 2 \cdot \frac{T}{\mu_V} \simeq 1.4 \, \frac{T}{\mu_V} \gg 1
\label{eq:v:ratio}
\eeqn
since $ T \gg \mu_V$ in the low-density QGP being created in the HIC experiments. Given the mentioned equivalence of relaxation times for axial ($\tau_A$) and helical ($\tau_H$) charges, $\tau_A \sim \tau_H$, and the large difference in their velocities~\eq{eq:v:ratio}, one can estimate that the wavelength of the propagating CVW will be much lower then the wavelength of the HVW. 
Therefore, the prospects of the HVW to exist in the QGP are substantially higher than the ones of the CVW since the HVW propagates faster.

We leave a more thorough analysis of the spectrum of waves allowed by the transport laws uncovered in this paper for a future analysis~\cite{ref:in:preparation}.

\subsection{Vortices in Dirac fluids}\label{sec:app:condmat}

In the condensed-matter context, the non-topological vortical structures may be created in finite samples of graphene~\cite{ref:LF}. These arguments are supported by a strong experimental evidence~\cite{ref:Geim}. 
However, the Dirac excitations in two-dimensional graphene do not posses helicity, while similar vortex (or, whirlpool) structures in three-dimensional systems are yet to be realized experimentally. The two- and three-dimensional Dirac materials are somewhat comparable, in particular, due to similarities in typical Fermi velocities, as well as in the electronic viscosities in the suspected hydrodynamic regimes.
Taking the vortex size $l \sim 1\,\mu\mathrm{m}$~\cite{ref:Geim} and the Fermi velocity $v_F \simeq c/300 \simeq 10^6\,\mathrm{m/s}$ one gets the vorticity in the terahertz region, $\omega \sim v_F/l \sim 10^{12} \, \mathrm{s}^{-1}$. 
Free Dirac excitations, on top of the rotating viscous electronic fluid, should thus generate the 
helical vortex excitation which propagates, at temperature $T= 100\,\mathrm{K}$, with the noticeable velocity $v_{\mathrm{\tiny{HVW}}} \sim 10^4 \, \mathrm{m/s}$. The existence of the helical vortical wave in this environment depends, as in the case of the quark-gluon plasma, on the helical relaxation time, the estimation of which is beyond the scope of this paper.

\section{Conclusions}\label{sec:conc}

In this paper, we exploited the invariance of the free Dirac Lagrangian under 
the $U(1)_H$ helical transformations to introduce the helicity charge 
current, $J_H^\mu$, and the corresponding helicity charge operator,
$\widehat{Q}_H$. The associated helical chemical potential, $\mu_H$, can 
be used to describe helical imbalance in thermal states. 
Physically, the helical charge (current) density can be considered as the difference between the axial charge (current) density of positive and negative energy eigenmodes.

We found that the helical degree of freedom participates in new transport phenomena, the helical vortical effects (HVE), which appear in a rigidly rotating ensemble of Dirac fermions at finite temperature and density. The HVE's generate currents of vector, axial, and helical charges and lead to new 
excitations propagating through the helical vortical matter (the helical vortical waves, HVW). Our estimations show that,
while the effects of the helical vortical waves are unlikely to be observable in quark-gluon plasma in high-energy physics, 
they could be probed in new Dirac materials in solid-state context. We suggested that the HVE may possibly be related to new triangle anomalies in quantum electrodynamics. 
In a thermodynamic context, the helical imbalance may lead to profound consequences for the chiral symmetry breaking in dense quark-gluon plasma~\cite{Chernodub:2020yaf}.

Based on a classical kinetic theory description of Fermi-Dirac 
particles with vector, axial and helical imbalance, we showed that the 
transport coefficients related to anomalous transport can be obtained 
thermodynamically as derivatives of the charge densities 
$Q_\ell = \partial P / \partial \mu_\ell$ with respect to the axial 
chemical potential $\mu_A$. Following the constraints imposed by the 
parities under the discrete $C$-, $P$- and $T$- inversions, we were 
able to express the higher-order corrections to the anomalous transport 
coefficients in terms of such thermodynamic functions. The exact 
proportionality coefficients are computed on the basis of thermal
field theory for matter under rotation.

We close this paper by remarking that the new effects uncovered herein 
do not rely on the persistence of helicity imbalance, which is 
expected to be suppressed due to quantum interactions.
Despite the fact that helicity is not conserved in realistic interacting gauge theories, we demonstrate that its relaxation time 
is of the same order as the chirality relaxation time, $\tau_H \simeq \tau_A$. Thus, both the helicity and the chirality should be treated on an equal footing. A relevant example of the application of the helicity current to polarization of hyperons in noncentral heavy-ion collisions is discussed in Ref.~\cite{Ambrus:2020oiw}.

\section*{Acknowledgements}
The authors thank A.~Cortijo, K.~Landsteiner, P.~Aasha and D.~Wagner for useful discussions.  The work of V.E.A. was supported by the Grants from the Romanian  National Authority for Scientific Research and Innovation, CNCS-UEFISCDI, having project numbers PN-III-P1-1.1-PD-2016-1423 and PN-III-P1-1.1-TE-2021-1707.

\appendix 

\section{Computational details} \label{app:comp}

This Appendix summarises the details regarding the computation 
of the thermal expectation values (t.e.v.s) discussed in 
Sec.~\ref{sec:QFT}. The computational technique was introduced 
in Ref.~\cite{ambrus14plb} and is presented in more detail 
in the book chapter \cite{ambrus19lnp}. 

This appendix is structured as follows. 
Subsection~\ref{app:comp:modes} presents details required for the derivation
of the expressions for the non-vanishing components of the charge currents and
stress-energy tensor, given in Eqs.~\eqref{eq:Jl_comps} and \eqref{eq:SET_comps} 
in the main text. A short mathematical interlude in Subsec.~\ref{app:comp:series}
summarises the procedure employed to derive analytical expressions in a 
power series with respect to the rotation parameter $\Omega$. 
The results relevant for the charge currents and stress-energy tensor
are derived in the Subsections~\ref{app:comp:Jl} and \ref{app:comp:SET}, 
respectively. All calculations presented in this section are performed 
with respect to the non-rotating (stationary) vacuum. 

\subsection{Modes and sesquilinear forms}\label{app:comp:modes}

For systems undergoing rigid rotation, it is convenient to consider a set of 
modes with respect to which the statistical operator $\hat{\varrho}$ 
in Eq.~\eqref{eq:rho} is diagonal. For this purpose, we employ the helicity eigenmodes $U_j$ derived in cylindrical coordinates $(t, \rho, \varphi, z)$ 
in Ref.~\cite{ambrus14plb}, which are summarised for the case of massless fermions 
below:
\begin{multline}
 U^\lambda_{p,k,m}(x) = \frac{e^{-i p t + i k z}}{4\pi}
 \begin{pmatrix}
 1 \\ 2\lambda
 \end{pmatrix} \\\otimes
 \begin{pmatrix}
  \sqrt{1 + \frac{2\lambda k}{p}} e^{i(m - \frac{1}{2})\varphi}
  J_{m -\frac{1}{2}}(q\rho) \\
  2i\lambda \sqrt{1 - \frac{2\lambda k}{p}} e^{i(m + \frac{1}{2})\varphi}
  J_{m +\frac{1}{2}}(q\rho)
 \end{pmatrix}.
 \label{eq:modes}
\end{multline}
In the above, $m = \pm \frac{1}{2}, \pm \frac{3}{2}, \dots$ is 
the eigenvalue of the total angular momentum operator $M^z$, while 
$\lambda = \pm \frac{1}{2}$ is the eigenvalue of the helicity operator $h$.
The eigenvalue of the linear momentum operator along the $z$ axis is $k$,
while $q = \sqrt{p^2 - k^2}$ is the magnitude of the transverse momentum.
The anti-particle modes are obtained via charge conjugation, 
$V_j = i \gamma^2 U_j^*$.

For the construction of rigidly-rotating thermal states, Iyer \cite{iyer82} argued 
that the modes with positive co-rotating energy $\widetilde{E}_j = E_j - \Omega m_j > 0$ 
should be interpreted as particle modes, while $E_j = \pm \lvert p_j \rvert$ is allowed to take
both positive and negative values. For brevity, in this paper we discuss 
only t.e.v.s computed with respect to the non-rotating (Minkowski) vacuum, keeping in mind
that the difference between the two approaches manifests itself as terms which depend
solely on the rotation parameter $\Omega$ (via the vorticity $\bm{\omega}$ or 
acceleration $\bm{a}$), being independent of the medium properties $T$ and $\mu_\ell$.
With respect to the Minkowski vacuum, the Minkowski energy $E_j = p_j$ is always positive. In this case, $\hat{\psi}$ can be decomposed 
as
\begin{align}
 \hat{\psi} (x) &= \sum_j [U_j(x) \hat{b}_j + V_j(x) \hat{d}^\dagger_j]\nonumber\\
 &= \sum_{\lambda = \pm 1/2} 
 \sum_{m = -\infty}^\infty \int_{0}^\infty dp
 \, p \int_{-p}^p dk \nonumber\\
 & \times [U^\lambda_{p,k,m}(x) \hat{b}^{\lambda}_{p,k,m} + 
 V^\lambda_{p,k,m}(x) \hat{d}^{\lambda;\dagger}_{p,k,m}].
 \label{eq:psi_M}
\end{align}

We now summarise the explicit expressions for the 
sesquilinear forms appearing in Eq.~\eqref{eq:sesquilinear}.
Due to charge conjugation, the sesquilinear forms for the 
anti-particle modes $V_j = i \gamma^2 U_j^*$ are related to those 
for the particle modes via
\begin{align}
 \mathcal{J}^\mu_{V/H}(V_j, V_j) 
 &= \mathcal{J}^\mu_{V/H}(U_j, U_j), \nonumber\\
 \mathcal{J}^\mu_{A}(V_j, V_j) &= -\mathcal{J}^\mu_A(U_j, U_j), \nonumber\\
 \mathcal{T}^{\mu\nu}(V_j, V_j) &= -\mathcal{T}^{\mu\nu}(U_j, U_j).
\end{align}
Furthermore, it can be seen that $\mathcal{J}^\mu_A(U_j,U_j) = 
\mathcal{J}^\mu_H(U_j,U_j) = 2\lambda_j \mathcal{J}^\mu_V(U_j, U_j)$, 
since $U_j$ is a simultaneous eigenvector of both $h$ and $\gamma^5$. 
The sesquilinear forms $\mathcal{J}^\mu_V(U_j, U_j)$
read \cite{Ambrus:2019ayb}:
\begin{align}
 \mathcal{J}^t_V(U_j, U_j) &= \frac{1}{8\pi^2} \left[J^+_{m_j}(q_j \rho) + 
 \frac{2\lambda_j k_j}{p_j} J^-_{m_j}(q_j \rho)\right],\nonumber\\
 \mathcal{J}^\varphi_V(U_j, U_j) &= \frac{q_j}{8\pi^2 p_j} J^\times_{m_j}(q_j \rho),\nonumber\\
 \mathcal{J}^z_V(U_j, U_j) &= \frac{1}{8\pi^2} \left[
 \frac{k_j}{p_j} J^+_{m_j}(q_j \rho) + 
 2\lambda_j J^-_{m_j}(q_j \rho)\right],
 \label{eq:app_JV_sesqui}
\end{align}
while $\mathcal{J}^\rho_{V/A/H}(U_j, U_j) = 0$.
The terms proportional to $k_j$ 
appearing in $\mathcal{J}^t_\ell(U_j, U_j)$ and $\mathcal{J}^z_\ell(U_j, U_j)$ were dropped in Eq.~\eqref{eq:Jl_comps} since they are odd 
with respect to $k_j \rightarrow -k_j$.

On the stress-energy tensor (SET) sector, we have \cite{Ambrus:2019ayb}
\begin{align}
 \mathcal{T}^{tt}(U_j, U_j) &= \frac{p_j}{8\pi^2} 
 \left[J_{m_j}^+(q_j\rho) + \frac{2\lambda_j k_j}{p_j} 
 J_{m_j}^-(q_j\rho)\right],\nonumber\\
 \mathcal{T}^{t\varphi}(U_j, U_j) &=
 \frac{1}{16\pi^2 \rho^2} 
 \left[\left(m_j - \frac{\lambda_j k_j}{p_j}\right) J_{m_j}^+(q_j \rho) \right. \nonumber\\
 & \hspace{-20pt} \left. +
 \left(\frac{2 \lambda_j k_j m_j}{p_j} - \frac{1}{2}\right) J_{m_j}^-(q_j \rho)\right] \nonumber\\
 & \hspace{-20pt} +\frac{q_j}{16\pi^2 \rho} J_{m_j}^\times(q_j\rho),\nonumber\\
 \mathcal{T}^{tz}(U_j, U_j) &= 
 \frac{\lambda_j (p^2_j + k_j^2)}{8\pi^2 p_j} J_{m_j}^-(q_j \rho)
 + \frac{k_j}{8\pi^2} J_{m_j}^+(q_j \rho),\nonumber\\
 \mathcal{T}^{\rho\rho}(U_j, U_j) &= \frac{q_j^2}{8\pi^2 p_j} 
 \left[J_{m_j}^+(q_j\rho) - \frac{m_j}{q_j \rho} 
 J_{m_j}^\times(q_j\rho)\right],\nonumber\\
 \mathcal{T}^{\varphi\varphi}(U_j, U_j) &= 
 \frac{q_j m_j}{8\pi^2 \rho^3 p_j} 
 J_{m_j}^\times(q_j\rho),\nonumber\\
 \mathcal{T}^{\varphi z}(U_j, U_j) &= 
 \frac{\lambda_j}{8\pi^2 \rho^2} 
 \left[m_j J_{m_j}^-(q_j\rho) - \frac{1}{2} J_{m_j}^+(q_j\rho)\right]
 \nonumber\\
 &\hspace{-20pt} + \frac{k_j}{16\pi^2 \rho^2 p_j} 
 \left[q_j \rho J_{m_j}^\times(q_j\rho) - 
 \frac{1}{2} J_{m_j}^-(q_j \rho) \right.\nonumber \\
 & \hspace{-20pt}\left. +
 m_j J_{m_j}^+(q_j \rho)\right],\nonumber\\
 \mathcal{T}^{zz}(U_j, U_j) &= 
 \frac{k_j^2}{8\pi^2 p_j} J_{m_j}^+(q_j\rho) + 
 \frac{\lambda_j k_j}{4\pi^2}  
 J_{m_j}^-(q_j\rho).
 \label{eq:app_SET_sesqui}
\end{align}
Upon substitution of the above results in Eq.~\eqref{eq:J_SET_modes_aux}, 
the terms which are odd with respect to $k_j \rightarrow -k_j$ can be dropped 
and Eq.~\eqref{eq:SET_comps} is reproduced. 

\subsection{Slow rotation limit}\label{app:comp:series}

In order to derive analytical expressions for the t.e.v.s 
$J^\mu_\ell \equiv \braket{:\widehat{J}^\mu_\ell:}$ and 
$T^{\mu\nu} \equiv \braket{:\widehat{T}^{\mu\nu}:}$, we consider 
the limit of slow rotation, in which the Fermi-Dirac factor 
$n_{\sigma,\lambda}(\tilde{p})$ introduced in Eq.~\eqref{eq:n_def} 
can be expanded as follows:
\begin{equation}
 n_{\sigma,\lambda}(\tilde{p}) = 
 \sum_{n = 0}^\infty \frac{(-\Omega m)^n}{n!}
 \frac{d^n}{dp^n} n_{\sigma,\lambda}(p).
 \label{eq:n_exp}
\end{equation}
Note that the tilde no longer appears over $p$ in the argument of $n_{\sigma,\lambda}$ 
on the right hand side. In order to illustrate the procedure, in 
what follows we take as an example the $t$ component of the 
charge current $J^t_\ell$. Substituting Eq.~\eqref{eq:n_exp} in Eq.~\eqref{eq:Jl_comps} gives
\begin{multline}
 J_\ell^t = \sum_{\sigma,\lambda} \frac{q^\ell_{\sigma,\lambda}}{4\pi^2} 
 \int_0^\infty dp\, p
 \sum_{n = 0}^\infty \frac{(-\Omega)^n}{n!} 
 \frac{\partial^n n_{\sigma,\lambda}(p)}{\partial p^n}\\ \times 
 \int_{0}^p dk
 \sum_{m = -\infty}^\infty
 m^n J_m^+(q\rho).
 \label{eq:app_Jt_n}
\end{multline}

The sum over $m$ typically involves terms which are quadratic in Bessel functions 
[see Eqs.~\eqref{eq:Jl_comps} and \eqref{eq:SET_comps}], which can be performed 
analytically using the following formulas~\cite{ambrus14plb}
\begin{align}
 \sum_{m = -\infty}^\infty m^{2n} J_m^+(z) =& \sum_{j = 0}^n 
 \frac{2 \Gamma(j + \frac{1}{2})}{j! \sqrt{\pi}} s_{n,j}^+ z^{2j},\nonumber\\
 \sum_{m = -\infty}^\infty m^{2n+1} J_m^-(z) =&
 \sum_{j = 0}^n \frac{2 \Gamma(j + \frac{3}{2})}{j! \sqrt{\pi}} s_{n,j}^+ z^{2j},\nonumber\\
 \sum_{m = -\infty}^\infty m^{2n+1} J_m^\times(z) =&
 \sum_{j = 0}^n \frac{2\Gamma(j + \frac{3}{2})}{(j+1)!\sqrt{\pi}} 
 s_{n,j}^+ z^{2j+1},
 \label{eq:summ}
\end{align}
where $n$ is a non-negative integer and it is understood that 
$\sum_m m^{2n+1} J_m^+ = \sum_m m^{2n} J^-_m = \sum_m m^{2n} J^\times_m = 0$.
The coefficient $s_{n,j}^+$ appearing above can be obtained from
\begin{equation}
 s_{n,j}^+ = \frac{1}{(2j+1)!} 
 \lim_{\alpha \rightarrow 0} 
 \frac{d^{2n+1}}{d\alpha^{2n+1}} 
 \left(2 \sinh\frac{\alpha}{2}\right)^{2j+1}
\end{equation}
and vanishes when $j > n$. For small values of 
$n - j \ge 0$, the first few coefficients are
\begin{align}
 s_{j,j}^+ &= 1, \quad s_{j+1,j} = \frac{1}{24}(2j+1)(2j+2)(2j+3),\nonumber\\
 s_{j+2,j}^+ &= \frac{1}{5760} (2j+1)(2j+2)(2j+3)
 \nonumber\\ & \times 
(2j+4)(2j+5)(10j+3).
 \label{eq:summ_s}
\end{align}
Performing the sum over $m$ using Eq.~\eqref{eq:summ} in 
Eq.~\eqref{eq:app_Jt_n} gives
\begin{multline}
 J_\ell^t = \sum_{\sigma,\lambda} \frac{q^\ell_{\sigma,\lambda}}{4\pi^2}  
 \int_0^\infty dp\, p
 \sum_{n = 0}^\infty \frac{(-\Omega)^n}{n!} 
 \frac{\partial^n n_{\sigma,\lambda}(p)}{\partial p^n} \\\times
 \sum_{j = 0}^n \frac{2 \Gamma(j + \frac{1}{2})}{j! \sqrt{\pi}} s_{n,j}^+ \rho^{2j}
 \int_{0}^p dk\, 
 q^{2j}.
 \label{eq:app_Jt_j}
\end{multline}
The integration with respect to $k$ of polynomials in $q$ can be performed using
\begin{equation}
 \int_{0}^p dk\, q^{2j} = \frac{j! \sqrt{\pi}}
 {2\Gamma(j + 3/2)} p^{2j + 1},
 \label{eq:intk}
\end{equation}
which in the case of Eq.~\eqref{eq:app_Jt_j} gives
\begin{multline}
 J_\ell^t = \sum_{\sigma,\lambda} \frac{q^\ell_{\sigma,\lambda}}{2\pi^2} 
 \int_0^\infty dp\, 
 \sum_{n = 0}^\infty \frac{(-\Omega)^n}{n!} 
 \frac{\partial^n n_{\sigma,\lambda}(p)}{\partial p^n}\\\times 
 \sum_{j = 0}^n \frac{s_{n,j}^+ \rho^{2j}}{2j+1} 
 \rho^{2j} p^{2j+2}.
 \label{eq:app_Jt_intk}
\end{multline}

Taking into account that the summation range of $j$ is between $0$ and $n$,
the sums with respect to $n$ and $j$ can be exchanged, such that 
$n$ now takes values from $j$ to $\infty$. Shifting down the 
summation end for $n$ from $j$ to $0$ gives in the case of Eq.~\eqref{eq:app_Jt_intk} 
\begin{multline}
 J_\ell^t = \sum_{n = 0}^\infty \Omega^{2n}
 \sum_{\sigma,\lambda} \frac{q^\ell_{\sigma,\lambda}}{4\pi^2} 
 \int_0^\infty dp\, 
 p^2 \frac{\partial^{2n} n_{\sigma,\lambda}(p)}{\partial p^{2n}}\\\times
 \sum_{j = 0}^\infty 
 \frac{(\rho\Omega)^{2j} (2j+2)!}{(2n+2j)!(2j+1)} s_{n+j,j}^+,
 \label{eq:app_Jt_intk_aux}
\end{multline}
where integration by parts was employed $2j$ times to get from 
$d^{2n+2j} n_{\sigma,\lambda}(p) / dp^{2n+2j}$ to 
$d^{2n} n_{\sigma,\lambda}(p) / dp^{2n}$. 
The integration with respect to $p$ can be performed by changing the 
derivatives of $n_{\sigma,\lambda}(p)$ with respect to $p$ to 
derivatives with respect to $\alpha_{\sigma,\lambda} \equiv \bm{q}_{\sigma,\lambda} \cdot \bm{\mu} / T$,
\begin{multline}
 \int_0^\infty dp\, p^r \frac{d^{2n} n_{\sigma,\lambda}}{dp^{2n}} \\
 = -r! \beta_0^{2n-r-1} \frac{d^{2n}}{d\alpha_{\sigma,\lambda}^{2n}} {\rm Li}_{r+1}(-e^{\alpha_{\sigma,\lambda}})\\
 = -\left(\frac{T}{\Gamma}\right)^{1+r-2n} r! {\rm Li}_{r+1-2n}(-e^{\alpha_{\sigma,\lambda}}).
 \label{eq:app_Li_intp}
\end{multline}
Applying the above to Eq.~\eqref{eq:app_Jt_intk_aux} gives
\begin{multline}
 J^t_\ell = -\sum_{n = 0}^\infty \left(\frac{\Gamma \Omega}{T}\right)^{2n}
 \sum_{\sigma,\lambda} \frac{q^\ell_{\sigma,\lambda} T^3}{2\pi^2} 
 {\rm Li}_{3-2n}(-e^{\alpha_{\sigma,\lambda}}) \\\times
 \sum_{j = 0}^\infty \frac{(\rho \Omega)^{2j} (2j+2)!}{\Gamma^3 (2n+2j)!(2j+1)} s_{n+j,j}^+.
 \label{eq:app_Jt_intp}
\end{multline}
The summation over $j$ can be performed in terms of the 
Lorentz factor $\Gamma$, vorticity $\bm{\omega}^2 = \Omega^2 \Gamma^4$ 
and acceleration $\bm{a}^2 = \Omega^2 \Gamma^2(\Gamma^2 - 1)$. For example, in the case $n = 0$ in Eq.~\eqref{eq:app_Jt_intp}, the summation yields
\begin{subequations}\label{eq:app_sumj}
\begin{align}
 \sum_{j = 0}^\infty \frac{(\rho \Omega)^{2j} (2j+2)!}{\Gamma^3 (2n+2j)!(2j+1)} s_{n+j,j}^+ =& 2\Gamma,
 \label{eq:app_sumj_n0}
\end{align}
while when $n = 1$, one obtains
\begin{multline}
 \Gamma^2 \Omega^2 \sum_{j = 0}^\infty \frac{(\rho \Omega)^{2j} (2j+2)!}{\Gamma^3 (2n+2j)!(2j+1)} s_{n+j,j}^+ \\
 = \frac{\Gamma}{12} (3\bm{\omega}^2 + \bm{a}^2).
 \label{eq:app_sumj_n1}
\end{multline}
\end{subequations}

The summation over $\sigma$ and $\lambda$ appearing in 
Eq.~\eqref{eq:app_Jt_intp} can be performed 
in terms of the thermodynamic quantities introduced in
Eq.~\eqref{eq:RKT_Li} when $r + 1 - 2n \ge 0$. This corresponds to 
the $n = 0$ and $n = 1$ terms in Eq.~\eqref{eq:app_Jt_intp}, which 
give
\begin{subequations}\label{app:Jt}
\begin{equation}
 J^t_\ell = \Gamma \left(Q_{\ell;{\rm cl}} + \frac{3\bm{\omega}^2 + \bm{a}^2}{2} \sigma^\tau_{\ell;{\rm cl}} \right) + J^{t,(h)}_\ell,
\end{equation}
where $J^{t,(h)}_{\ell}$ is a high-order correction given by
\begin{multline}
 J^{t,(h)}_{\ell} = -\frac{\Omega^4 \Gamma}{T}\\\times 
 \sum_{n = 0}^\infty \left(\frac{\Gamma \Omega}{T}\right)^{2n}
 \sum_{\sigma,\lambda} \frac{q^\ell_{\sigma,\lambda}}{2\pi^2} 
 {\rm Li}_{-1-2n}(-e^{\alpha_{\sigma,\lambda}}) \\\times
 \sum_{j = 0}^\infty \frac{(\rho \Omega)^{2j} (2j+2)!}{(2n+2j+4)!(2j+1)} s_{n+j+2,j}^+.
 \label{eq:app_Jt_h}
\end{multline}
\end{subequations}
In what follows, we will label the contributions to thermal 
expectation values involving polylogarithms of negative index 
using the $(h)$ superscript, as above.
To understand the properties of such terms, one may look at the 
large temperature behaviour of the polylogarithm,
\begin{equation}
 {\rm Li}_n(-e^{\alpha_{\sigma,\lambda}}) = 
 - \sum_{s = 0}^\infty
 (1 - 2^{1 + s - n}) \frac{\zeta(n - s)}{s!} \alpha_{\sigma,\lambda}^s.
\end{equation}
Summing now over $\sigma$ and $\lambda$, we obtain
\begin{equation}
 \sum_{\sigma,\lambda} {\rm Li}_n(-e^{\alpha_{\sigma,\lambda}})
 = - \sum_{s = 0}^\infty
 (1 - 2^{1 + s - n}) \frac{\zeta(n - s)}{s!} \mathcal{S}_s,
 \label{eq:app_polsum_Li}
\end{equation}
where $\mathcal{S}_s = \sum_{\sigma,\lambda} \alpha_{\sigma,\lambda}^s$ is
given explicitly by
\begin{multline}
 \mathcal{S}_s = \frac{1}{T^s}[
 (\mu_V + \mu_A + \mu_H)^s +    
 (\mu_V - \mu_A - \mu_H)^s \\
 + (-\mu_V + \mu_A - \mu_H)^s + 
 (-\mu_V - \mu_A + \mu_H)^s].
\end{multline}
When $s = 1$, the sum vanishes, while for odd values of $s$ higher than $1$, 
it is not difficult to see that the result must be of order $s$ with respect to 
$\mu_V$, $\mu_A$, $\mu_H$ and their products. Explicitly, we have
\begin{align}
 \mathcal{S}_0 &= 4, \quad
 \mathcal{S}_1 = 0, \quad 
 \mathcal{S}_2 = \frac{4 \bm{\mu}^2}{T^2},\quad
 \mathcal{S}_3 = \frac{24 \mu_V \mu_A \mu_H}{T^3}, \nonumber\\
 \mathcal{S}_4 &= 
 \frac{4}{T^4}[\mu_V^4 + \mu_A^4 + \mu_H^4 \nonumber\\
 &+ 6(\mu_V^2 \mu_A^2 + \mu_V^2 \mu_H^2 + \mu_A^2 \mu_H^2)], \nonumber\\
 \mathcal{S}_5 &= 
 \frac{80 \mu_V \mu_A \mu_H \bm{\mu}^2}{T^5}.
\end{align}
Charge prefactors $q^{\ell}_{\sigma,\lambda}$ can 
be obtained, e.g., by taking derivatives of Eq.~\eqref{eq:app_polsum_Li} 
with respect to $\mu_\ell$:
\begin{multline}
 \sum_{\sigma,\lambda} q^{\ell_1}_{\sigma,\lambda} 
 \cdots q^{\ell_r}_{\sigma,\lambda} 
 {\rm Li}_n(-e^{\alpha_{\sigma,\lambda}}) = 
 T^r \\\times 
 \frac{\partial^r}{\partial \mu_{\ell_1} \cdots \partial \mu_{\ell_r}}
 \sum_{\sigma,\lambda} {\rm Li}_{n + r}(-e^{\alpha_{\sigma,\lambda}}).
\end{multline}
In the case of $J^{t,(h)}_\ell$, the leading order contribution is proportional to
\begin{multline}
 \frac{\Omega^4}{T} \sum_{\sigma,\lambda} q^\ell_{\sigma,\lambda} 
 {\rm Li}_{-1}(-e^{\alpha_{\sigma,\lambda}}) =
 \frac{15 \Omega^4 \zeta(-3)}{3!} 
 \frac{\partial \mathcal{S}_3}{\partial \mu_\ell} + \dots \\
 = 
 \frac{\mu_V \mu_A \mu_H \Omega^4}{2\mu_\ell T^3} + O(T^{-5}).
\end{multline}
There are no lower order contributions since
$\partial \mathcal{S}_0 / \partial \mu_\ell = 0$, 
$\mathcal{S}_1 = 0$ and $\zeta(-2) =0$.
Thus, we conclude that 
$J^{t,(h)}_\ell = O(\mu_V \mu_A \mu_H \Omega^4 / \mu_\ell T^3)$.

\subsection{V/A/H charge currents}\label{app:comp:Jl}

We now apply the strategy outlined in Subsec.~\ref{app:comp:series} 
for the case of the charge current $J^\mu_\ell$. We present 
as an intermediate step the relation similar to 
that in Eq.~\eqref{eq:app_Jt_intk_aux} in the case of the non-vanishing 
components of $J^\mu_\ell$:
\begin{multline}
 \begin{pmatrix}
  J^t_\ell \\
  J^\varphi_\ell \\
  J^z_\ell
 \end{pmatrix} = 
 \sum_{n = 0}^\infty \Omega
 \sum_{\sigma,\lambda} 
 \frac{q^\ell_{\sigma,\lambda}}{4\pi^2}
 \int_0^\infty dp\, p^2
 \frac{d^{2n} n_{\sigma,\lambda}}{dp^{2n}} \\\times
 \sum_{j=  0}^\infty 
 \frac{s_{n+j,j}^+ (\rho \Omega)^{2j} (2j+2)!}{(2n+2j+1)!(2j+1)}
 \begin{pmatrix}
  2n+2j+1\\
  \Omega(2j+1)\\
  2\lambda \Omega p^{-1} (2j+1)
 \end{pmatrix}.
 \label{eq:Jl_comps_aux}
\end{multline}
Starting from 
the decomposition in Eq.~\eqref{eq:currents}, the charge 
density $Q_\ell = u_\mu J^\mu_\ell$, circular conductivity
$\sigma^\tau_\ell = -\tau_\mu J^\mu_\ell / \tau^2$ and 
vortical conductivity 
$\sigma^\omega_\ell = -\omega_\mu J^\mu_\ell / \omega^2$ can 
be calculated via
\begin{gather}
 Q_\ell = \Gamma(J_\ell^t - \rho^2 \Omega J_\ell^\varphi), \nonumber\\
 \sigma^\tau_\ell = \frac{\Omega J_\ell^t - J_\ell^\varphi}{\Omega^3 \Gamma^3}, \qquad 
 \sigma^\omega_\ell = \frac{J^z}{\Omega \Gamma^2}.
\end{gather}
Using Eqs.~\eqref{eq:Jl_comps_aux}, we find
\begin{multline}
 \begin{pmatrix}
  Q_\ell \\ \sigma^\tau_\ell \\ \sigma^\omega_\ell
 \end{pmatrix}
  = \sum_{n = 0}^\infty \Omega^{2n}
 \sum_{\sigma,\lambda} 
 \frac{q^\ell_{\sigma,\lambda}}{4\pi^2}
 \int_0^\infty dp\, p^2
 \frac{d^{2n} n_{\sigma,\lambda}}{dp^{2n}}
  \\\times  \sum_{j=  0}^\infty 
 \frac{s_{n+j,j}^+ (\rho\Omega)^{2j} (2j+2)!}
 {(2n+2j+1)!(2j+1)}
 \begin{pmatrix}
  2n\Gamma^2 + 2j+1 \\ 2n / \Omega^2 \Gamma^2 \\ q^A_{\sigma,\lambda} (2j+1) /  p\Gamma
 \end{pmatrix},
 \label{eq:Jl_coeffs_sum}
\end{multline}
where $q^A_{\sigma,\lambda} = 2\lambda$ was used in the expression for
$\sigma^\omega_\ell$. After performing the $p$ integral using 
Eq.~\eqref{eq:app_Li_intp} and the summation with respect to $j$ 
as indicated in Eq.~\eqref{eq:app_sumj}, it can be seen that the 
$n = 0$ contribution to $Q_\ell$ coincides 
with $Q_{\ell;{\rm cl}}$ introduced in Eq.~\eqref{eq:RKT_Li}. 
For $n > 0$, quantum corrections are uncovered:
\begin{subequations}\label{eq:app_Ql}
\begin{equation}
 Q_\ell = Q_{\ell;{\rm cl}} + \frac{3(\bm{\omega}^2 + \bm{a}^2)}{2} \sigma^\tau_{\ell;{\rm cl}} + Q^{(h)}_\ell,
 \label{app:app_Ql_full}
\end{equation}
where $Q^{(h)}_\ell$ is subleading with respect to the temperature,
taking the form
\begin{multline}
 Q^{(h)}_\ell = -\frac{\Omega^4}{2\pi^2 T}\\\times 
 \sum_{n = 0}^\infty \left(\frac{\Omega \Gamma}{T}\right)^{2n}
 \sum_{\sigma,\lambda} 
 q^\ell_{\sigma,\lambda} {\rm Li}_{-1-2n}(-e^{\alpha_{\sigma,\lambda}}) \\
 \times \sum_{j = 0}^\infty \frac{s^+_{j+n+2,j} (\rho \Omega)^{2j} (2j+2)!}{(2n+2j+5)!(2j+1)} [(2n+4)\Gamma^4 + 2j + 1] \\
 = O\left(\frac{\Omega^4 \mu_V \mu_A \mu_H}{\mu_\ell T^3}\right).
 \label{eq:app_Ql_h}
\end{multline}
\end{subequations}
The leading order contribution to $Q^{(h)}_\ell$
comes from the $n = 0$ term, while terms at higher $n$ are penalised due 
to the $T^{-2n}$ factors.

As can be seen in Eq.~\eqref{eq:Jl_coeffs_sum}, 
the circular conductivity receives its leading order contribution from 
the $n = 1$ term and is in fact equal to the classical 
term $\sigma^\tau_{\ell;{\rm cl}}$ introduced in Eq.~\eqref{eq:RKT_Li},
\begin{subequations}\label{eq:app_sigmat_l}
\begin{equation}
 \sigma^\tau_\ell = \sigma^\tau_{\ell;{\rm cl}} + 
 \sigma^{\tau,(h)}_\ell,
 \label{eq:app_sigmat_l_full}
\end{equation}
where the quantum correction $\sigma^{\tau,(h)}_\ell$,
\begin{multline}
 \sigma^{\tau,(h)}_\ell = -\frac{\Omega^2}{2\pi^2 \Gamma^2 T} \\\times 
 \sum_{n = 0}^\infty
 \left(\frac{\Omega \Gamma}{T}\right)^{2n}
 \sum_{\sigma,\lambda} q^\ell_{\sigma,\lambda}
    {\rm Li}_{-1-2n}(-e^{\alpha_{\sigma,\lambda}})
    \\ \times \sum_{j = 0}^\infty 
 \frac{s^+_{j+n+2,j} (\rho \Omega)^{2j} (2j+2)!}{(2n+2j+5)!(2j+1)}
 (2n+4) \\
 = O\left(\frac{\Omega^2 \mu_V \mu_A \mu_H}{\mu_\ell T^3}\right),
 \label{eq:app_sigmat_l_h}
\end{multline}
\end{subequations}
exhibits the same structure and same leading order contribution as the correction
$Q^{(h)}_\ell$. 

The computation of the vortical conductivity involves the product
$q^A_{\sigma,\lambda} q^\ell_{\sigma,\lambda}$, discussed in 
Eq.~\eqref{eq:qAql}. Using the expressions for $\sigma^\omega_{\ell;{\rm cl}}$
and $\sigma^\Pi_{\ell;{\rm cl}}$ introduced in Eq.~\eqref{eq:RKT_Li}, 
it can be seen that
\begin{align}
 \sigma^\omega_\ell &= \sigma^\omega_{\ell;{\rm cl}} +
 \frac{\bm{\omega}^2+ 3\bm{a}^2}{24} 
 \sigma^\Pi_{\ell;{\rm cl}} + \sigma^{\omega,(h)}_\ell,\nonumber\\
 \sigma^{\omega,(h)}_\ell &= 
 -\frac{\Omega^4}{T^2} \nonumber\\
 & \times \sum_{n = 0}^\infty 
 \left(\frac{\Omega \Gamma}{T}\right)^{2n} \sum_{\sigma,\lambda} 
 \frac{q^A_{\sigma,\lambda} q^\ell_{\sigma,\lambda}}{4 \pi^2 }
 {\rm Li}_{-2-2n} (-e^{\alpha_{\sigma,\lambda}}) \nonumber\\
 & \times \sum_{j = 0}^\infty \frac{s^+_{j+n+2,j} (\rho\Omega)^{2j}(2j+2)!}{(2j+n+5)!}.
 \label{eq:app_sigmao}
\end{align}
The quantum correction proportional to $\sigma^\Pi_{\ell;{\rm cl}}$ 
makes a $T,\mu_\ell$-independent contribution to $\sigma^\omega_A$,
when $\sigma^\Pi_{A;{\rm cl}} = 1/\pi^2 + O(T^{-3})$, as indicated 
in Eq.~\eqref{eq:RKT_sigmaPil}. 
This contribution can be regarded as a vacuum term which appears due to the 
fact that the thermal expectation values discussed in this appendix 
are computed with respect to the static Minkowski vacuum. If the
calculation is performed with respect to the rotating vacuum proposed by 
Iyer \cite{iyer82}, this term no longer appears, i.e. 
$(\sigma^\omega_A)_{\rm rot} = \sigma^\omega_A - \lim_{T,\mu_\ell \rightarrow 0} (\sigma^\omega_A)$.
No such contributions can be found for $\sigma^\omega_{V/H}$.
The correction $\sigma^{\omega,(h)}_\ell$ behaves differently 
for $\ell = A$ than when $\ell = V/H$. Specifically, we find
$\sigma^{\omega,(h)}_A = O(\Omega^4 \mu_V\mu_A\mu_H/ T^5)$, while 
$\sigma^{\omega,(h)}_{V/H} = O(\Omega^4 \mu_{H/V} / T^3)$.

The results reported in this section are consistent with results 
previously reported in the literature. Our results correspond to the 
case of free massless fermions at finite $\mu_V$, $\mu_A$ and $\mu_H$.
In Sections 6 and 7 of Ref.~\cite{Ambrus:2019ayb}, similar results are
obtained in the case of vanishing axial chemical potential ($\mu_A = 0$), but 
at finite fermion mass. 
The case of finite $\mu_A$ but vanishing $\mu_H$ was considered 
in Ref.~\cite{buzzegoli18} (see Tables~1 and 2 therein). 

\subsection{Stress-energy tensor}\label{app:comp:SET}

We now discuss the properties of the stress-energy tensor. The pressure $P$,
circular and vortical heat conductivities $\sigma^\tau_\varepsilon$ and 
$\sigma^\omega_\varepsilon$ and shear stress coefficients $\Pi_1$ and $\Pi_2$ 
can be obtained via
\begin{align}
 P =& \frac{\Gamma^2}{3}(T^{tt} - 2\rho^2 \Omega T^{t\varphi} + \rho^4 \Omega^2 T^{\varphi\varphi}), \nonumber\\
 \sigma^\omega_\varepsilon =& \frac{1}{\Omega \Gamma} (T^{tz} - 
 \rho^2 \Omega T^{\varphi z}), \nonumber\\
 \sigma^\tau_\varepsilon =& \frac{1}{\Omega^3 \Gamma^2} \left[\Omega T^{tt} + 
 \rho^2 \Omega T^{\varphi \varphi} - (1 + \rho^2 \Omega^2) T^{t\varphi}\right],\nonumber\\
 \Pi_1 =& \frac{2(P - T^{zz})}{\rho^2 \Omega^6 \Gamma^8}, \qquad 
 \Pi_2 = \frac{\Omega T^{tz} - T^{\varphi z}}{\Omega^4 \Gamma^5}.
\end{align}
Based on the SET components displayed in Eq.~\eqref{eq:SET_comps}, the formalism 
from the previous subsection can be applied to derive the pressure,
\begin{subequations}\label{eq:app_P}
\begin{multline}
 P = P_{\rm cl} + \frac{\bm{a}^2 + 3\bm{\omega}^2}{12} \sigma^\omega_{A;{\rm cl}} \\ 
 + \frac{45\bm{\omega}^4 + 46\bm{\omega}^2 \bm{a}^2 - 51\bm{a}^4}{8640} \sigma^\Pi_{A;{\rm cl}} + P^{(h)},
\end{multline}
where $P_{\rm cl}$, $\sigma^\omega_{A;{\rm cl}}$ and $\sigma^\Pi_{A; {\rm cl}}$ can 
be found in Eq.~\eqref{eq:RKT_Li} and
\begin{multline}
 P^{(h)} = -\frac{\Omega^6 \Gamma^2}{6\pi^2 T^2} \sum_{n = 0}^\infty \left(\frac{\Omega \Gamma}{T}\right)^{2n} \\\times
 \sum_{\sigma,\lambda} 
 {\rm Li}_{-2-2n}(-e^{\alpha_{\sigma,\lambda}})
 \sum_{j = 0}^\infty \frac{(\rho \Omega)^{2j}(2j+2)!}{(2n+2j+7)!} \Bigg[\\
 4s^+_{j+n+3,j} \left(\frac{4j+5}{2j+1} + 2n + \frac{2j+1-2n}{4\Gamma^2}\right) \\
 - \rho^2 \Omega^2 (j+1) (j+2)(2j+3) s^+_{j+n+3,j+1}
 \Bigg] \\
 = O\left(\frac{\Omega^6 \mu_V \mu_A \mu_H}{T^5}\right).
\end{multline}
\end{subequations}

The heat vortical conductivity $\sigma^\omega_\varepsilon$ is given by
\begin{align}
 \sigma^\omega_\varepsilon &= Q_{A;{\rm cl}} + 
 \frac{\bm{\omega}^2 + \bm{a}^2}{2} \sigma^\tau_{A;{\rm cl}} + 
 \sigma^{\omega,(h)}_\varepsilon,\nonumber\\
 \sigma^{\omega,(h)}_\varepsilon &= -\frac{\Omega^4}{2\pi^2 T} \nonumber\\ 
 & \times \sum_{n = 0}^\infty \left(\frac{\Omega \Gamma}{T}\right)^{2n} \sum_{\sigma,\lambda} 
 q^A_{\sigma,\lambda} {\rm Li}_{-1-2n}(-e^{\alpha_{\sigma,\lambda}})\nonumber\\
 & \times \sum_{j = 0}^\infty \frac{s^+_{j+n+2,j} (\rho \Omega)^{2j} 
 (2j+2)! (1-n j)}{(2n+2j+5)!(2j+1)} \nonumber\\
 =& O\left(\frac{\Omega^4 \mu_V \mu_H}{T^3}\right),
 \label{eq:app_sigmao_eps}
\end{align}
where $Q_{A;{\rm cl}}$ and $\sigma^\tau_{A;{\rm cl}}$ can be found 
in Eq.~\eqref{eq:RKT_Li}.

The heat circular conductivity $\sigma^\tau_\varepsilon$ is given by
\begin{subequations}\label{eq:app_sigmat_eps}
\begin{equation}
 \sigma^\tau_\varepsilon = -\frac{1}{3} \sigma^\omega_{A;{\rm cl}} - 
 \frac{31\bm{a}^2 + 39 \bm{\omega}^2}{360} \sigma^\Pi_{A;{\rm cl}} + \sigma^{\tau,(h)}_\varepsilon,
\end{equation}
where $\sigma^\omega_{A;{\rm cl}}$ and $\sigma^\Pi_{A;{\rm cl}}$ 
were introduced in Eq.~\eqref{eq:RKT_Li}. The higher-order term $\sigma^{\tau,(h)}_\varepsilon$ is given by
\begin{multline}
 \sigma^{\tau,(h)}_\varepsilon = -\frac{\Omega^4}{2\pi^2 T^2} \\ \times \sum_{n = 0}^\infty \left(\frac{\Omega \Gamma}{T}\right)^{2n} \sum_{\sigma,\lambda} 
 {\rm Li}_{-2-2n}(-e^{\alpha_{\sigma,\lambda}})\\
 \times \sum_{j = 0}^\infty \frac{(\rho \Omega)^{2j} (2j+2)!}{(2n+2j+7)!} \Bigg\{\\
 8s^+_{j+n+3,j} \left[\frac{n(j+1)+1}{2j+1} + 1 + \frac{j+2}{4\Gamma^2}\right]\\
 - \frac{1+\rho^2 \Omega^2}{8} (2j+2)(2j+3)(2j+4) s^+_{n+j+3,j+1}
 \Bigg\} \\
 = O\left(\frac{\Omega^4 \mu_V \mu_A \mu_H}{T^5}\right).
\end{multline}
\end{subequations}

The shear stress coefficient $\Pi_1$ is given by
\begin{subequations}\label{eq:app_Pi1}
\begin{equation}
 \Pi_1 = -\frac{2}{27} \sigma^\Pi_{A;{\rm cl}} + \Pi^{(h)}_1,
\end{equation}
where $\sigma^\Pi_{A;{\rm cl}}$ is given in Eq.~\eqref{eq:RKT_Li} and
\begin{multline}
 \Pi^{(h)}_1 = -\frac{\Omega^2}{3\pi^2 T^2 \Gamma^6} \\\times \sum_{n = 0}^\infty \left(\frac{\Omega \Gamma}{T}\right)^{2n} \sum_{\sigma,\lambda} 
 {\rm Li}_{-2-2n}(-e^{\alpha_{\sigma,\lambda}}) \\\times
 \sum_{j = 0}^\infty (\rho \Omega)^{2j}(2j+4)! 
 \Bigg[
 \frac{s^+_{n+j+4,j+1}}{(2n+2j+9)!}\\ \times 
 \left(4n + 5 - \frac{2n+6}{2j+3} + 2n + \frac{2j+3-2n}{\Gamma^2}\right) \\ 
 - \frac{(j+1) s^+_{j+n+3,j+1}}{2(2n+2j+7)!} \Bigg]
 = O\left(\frac{\Omega^2 \mu_V \mu_A \mu_H}{T^5}\right).
\end{multline}
\end{subequations}

Finally, the coefficient $\Pi_2$ can be obtained as
\begin{subequations}\label{eq:app_Pi2}
\begin{equation}
 \Pi_2 = -2 \sigma^\tau_{A;{\rm cl}} + \Pi^{(h)}_2,
\end{equation}
where $\sigma^\tau_{A;{\rm cl}}$ can be read from Eq.~\eqref{eq:RKT_Li}, while
\begin{multline}
 \Pi^{(h)}_2 = -\frac{\Omega^2}{8\pi^2 \Gamma^4 T} \\\times 
 \sum_{n = 0}^\infty \left(\frac{\Omega \Gamma}{T}\right)^{2n} \sum_{\sigma,\lambda} q^A_{\sigma,\lambda}
 {\rm Li}_{-1-2n}(-e^{\alpha_{\sigma,\lambda}})\\\times
 \sum_{j = 0}^\infty \frac{(\rho \Omega)^{2j}(2j+4)!}{(2n+2j+5)!(2j+3)} 
 \Big(s^+_{n+j+2,j}  \\ 
 - \frac{j+1}{n+j+3} s^+_{n+j+3,j+1}\Big)
 = O\left(\frac{\Omega^2 \mu_V \mu_H}{T^3}\right).
\end{multline}
\end{subequations}

As in the case of the charge currents, the results 
reported here for the stress-energy tensor are in perfect agreement 
with the massless limit of those found in Section 8 of Ref.~\cite{Ambrus:2019ayb} for helical fermions at vanishing $\mu_A$, 
as well as with those reported in Tables~1 and 2 of Ref.~\cite{buzzegoli18}
for chiral fermions at vanishing $\mu_H$.

\section{Evaluation of the helicity relaxation time}\label{app:tauH}

In this appendix, we provide the details for the evaluation of the helicity relaxation time, based on the integral in Eq.~\eqref{eq:tauH_inv}, which we rewrite as
\begin{equation}
 \tau_H^{-1} = \frac{8}{3} (2\pi)^6 g \alpha_{\rm QCD}^2 \beta^3 \int dP dK f_{0\bp} f_{0\bk} I_\Omega,
 \label{eq:app_tauH_1}
\end{equation}
where 
\begin{multline}
 I_\Omega = \int
 dP' dK' (1 - \cos\theta_{cm})^2
 \delta^4(p + k - p' -k')  
 \\\times 
 \tilde{f}_{0\bp'} \tilde{f}_{0\bk'}(\tilde{f}_{0\bp} + f_{0\bp'}).
\end{multline}

The integral $I_\Omega$ introduced above can be evaluated in the center of mass (CM) frame, where 
\begin{equation}
 \delta^4(p + k - p' - k') \rightarrow \delta(E_{cm} - p'{}^0 - k'{}^0) \delta(-\bm{p}' - \bm{k}'),
\end{equation}
where $E_{cm}^2 = (p + k)^2$.
The above delta function allows the $dK'$ integral to be performed, leading to the replacement $(k'{}^0, \bm{k}')= (p'{}^0, -\bm{p}')$ inside the integrand. We further write $dP' = p'{}^0 dp'{}^0 d\Omega_{p'} / (2\pi)^3$ and perform the $p'{}^0$ integral, which amounts to replacing $k'{}^0,p'{}^0 \rightarrow E_{cm}/2$ everywhere.

The boost to the CM frame changes the arguments of the FD distributions that depend on $\bm{p}'$ and $\bm{k}'$. The exponential $e^{\beta p'{}^0} \equiv e^{\beta p' \cdot u_l}$ involves the four-product between the momentum $p'{}^\mu$ and the four-velocity $u^\mu_l = (1,0,0,0)$ of the plasma in the fluid rest frame.  In the center of mass frame, we have 
\begin{equation}
 u^\mu_l \rightarrow u^\mu_{cm} = \frac{1}{E_{\rm cm}} (p^0 + k^0, -\bp - \bk)^\mu.
\end{equation}
Taking into account also the effect of the delta function discussed above, the arguments of the equilibrium distributions will change according to 
\begin{align}
 p' \cdot u &\rightarrow \frac{1}{2}(p^0 + k^0 + x \lvert \bp+\bk \rvert), \nonumber\\
 k' \cdot u &\rightarrow \frac{1}{2}(p^0 + k^0 - x \lvert \bp+\bk \rvert),
\end{align}
where $x = 2\bp' \cdot (\bp + \bk) / (E_{cm} \lvert \bp + \bk \rvert)$ represents the cosine of the angle between $\bp'$ (in the CM frame) and $\bp + \bk$ (in the laboratory frame).

Finally, the factor $(1 - \cos\theta_{cm})^2$ related to the angle between $\bm{p}'$ and $\bm{p}$ measured in the CM frame can be computed in a Lorentz-covariant way via
\begin{equation}
1 - \cos\theta_{cm} = \frac{4 p \cdot p'}{E_{cm}^2}.
\end{equation}
In order to evaluate the four-product $p \cdot p'$, we must boost $p$ into the center of mass frame:
\begin{equation}
 p^\mu \rightarrow p^\mu_{cm} = (E_{cm}/2, \bp_{cm}),
\end{equation}
where $\bp_{cm}$ is given by
\begin{equation}
 \bp_{cm} = \bp - \frac{p^0 + E_{cm}/2}{p^0 + k^0 + E_{cm}} (\bp + \bk).
\end{equation}

Going now back to Eq.~\eqref{eq:app_tauH_1}, we have 
\begin{multline}
 I_\Omega = \frac{1}{(2\pi)^5} \int_{-1}^1 dx\, \tilde{f}_{0\bp'} \tilde{f}_{0\bk'} (\tilde{f}_{0\bp} + f_{0\bp'}) \\\times
 \int_0^{2\pi} \frac{d\varphi_{p'}}{2\pi} (1 - \cos\theta_{cm})^2.
\end{multline}
In the above, $x = \cos\theta_{p'}$ and $\theta_{p'}$ is the angle between $\bp'$ and $(\bp + \bk)$, while $\varphi_{p'}$ measures an angle in the plane perpendicular to $(\bp + \bk)$. Writing now $\bp_{cm} = \bp_{\lvert \rvert} + \bp_{\perp}$, where 
\begin{equation}
 p_{\lvert \rvert} = \frac{\bp_{cm} \cdot (\bp + \bk)}{\lvert \bp + \bk \rvert} = \frac{E_{cm}}{\lvert \bp + \bk \rvert} \left(\frac{p^0}{2} - k^0\right)
\end{equation}
and $p_\perp = (p_{cm}^2 - p^2_{\lvert \rvert})^{1/2}$ with $p_{cm} = E_{cm} / 2$, the $\varphi_{p'}$ integral can be performed as 
\begin{multline}
 \int_0^{2\pi} \frac{d\varphi_{p'}}{2\pi} (1 - \cos\theta_{cm})^2 = \left(1- \frac{p_{\lvert \rvert}}{p_{cm}}x \right)^2 \\
 + \frac{1 - x^2}{2} \frac{p_\perp^2}{p_{cm}^2}.
\end{multline}

We now parametrize $\bk$ in terms of its length $k$, $\cos \gamma = \bk \cdot \bp / (k p)$ and the azimuthal angle $\varphi_k$. With this parametrization, the integrand in Eq.~\eqref{eq:app_tauH_1} becomes independent of the angular parametrization of $\bp$:
\begin{multline}
 \tau_H^{-1} = \frac{2 g \alpha_{\rm QCD}^2 \beta^3}{3 \pi^3} \int_{-1}^1 d\cos\gamma \int_0^\infty dp\, p \\
 \times \int_0^\infty dk\, k \int_{-1}^1 dx\, f_{0\bp} f_{0\bk} \tilde{f}_{0\bp'} \tilde{f}_{0\bk'} (\tilde{f}_{0\bp} + f_{0\bp'}) \\\times 
 \left[\left(1- \frac{p_{\lvert \rvert}}{p_{cm}}x \right)^2 + \frac{1 - x^2}{2} \frac{p_\perp^2}{p_{cm}^2}\right].
\end{multline}
We now isolate the momentum magnitude by introducing the coordinates 
\begin{equation}
 z = \beta(p + k), \quad 
 \delta = \frac{p - k}{p + k},
\end{equation}
such that $0 \le z < \infty$ and $-1 \le \delta \le 1$. With this parametrization, the expressions appearing in the distribution functions are modified to
\begin{align}
 \beta p^0 &\rightarrow \frac{z}{2} (1 + \delta), &
 \beta p'{}^0 &\rightarrow \frac{z}{2} (1 + x \xi),\nonumber\\
 \beta k^0 &\rightarrow \frac{z}{2} (1 - \delta), &
 \beta k'{}^0 &\rightarrow \frac{z}{2} (1 - x\xi),
\end{align}
where we introduced the variable $\xi$ via 
\begin{equation}
 \xi = \sqrt{1 - \frac{1 - \delta^2}{2} (1 - \cos\gamma)},
\end{equation}
such that $d\xi = \frac{1 -\delta^2}{4\xi} d\cos\gamma$. We then have
\begin{equation}
\tau_H = \frac{3\pi^3 \beta}{g\alpha_{\rm QCD}^2 \mathcal{I}},\label{eq:app_tauH_res}
\end{equation}
where 
\begin{multline}
 \mathcal{I} = \int_0^\infty dz\, z^3 \int_{-1}^1 dx \int_{-1}^1 d\delta \int_{\lvert \delta \rvert}^1 d\xi \\\times \left[\frac{3 - x^2}{2} \xi + x(1 - 3\delta) + \frac{3x^2 - 1}{8\xi}(1 - 3\delta)^2\right] \\\times
 [e^{\frac{z}{2}(1+\delta)}+1]^{-1}
 [e^{\frac{z}{2}(1-\delta)}+1]^{-1} \\\times 
 [e^{-\frac{z}{2}(1+x \xi)}+1]^{-1}
 [e^{\frac{z}{2}(1 - x\xi)}+1]^{-1} \\\times
 \{
 [e^{-\frac{z}{2}(1+\delta)}+1]^{-1} + [e^{\frac{z}{2}(1+x\xi)}+1]^{-1}\}.
 \label{eq:app_tauH_I}
\end{multline}
While an exact analytical closed form result for the above integral is missing, we find numerically 
\begin{equation}
\mathcal{I} \simeq 4.81255.
\end{equation}
Substituting the above result into Eq.~\eqref{eq:app_tauH_res}, the typical value of the helicity relaxation time can be obtained as shown in Eq.~\eqref{eq:tauH_estimate} of the main text.



\begin{thebibliography}{99}
\bibitem{footnote:2}
We leave beside the scope of our work the conformal invariance which reflects a very different kind of symmetry of the theory of massless particles~\eq{eq:L}.
\end{thebibliography}

\end{document}